\documentclass[useAMS,usenatbib]{mn2e}
\usepackage[english]{babel}

\usepackage{fancyhdr}
\usepackage{amsfonts}
\usepackage{amsmath}
\usepackage{amssymb}
\usepackage{multicol}
\usepackage{layout}
\usepackage{graphicx}
\usepackage{subfigure}
\usepackage{multirow}

\usepackage{times}
\usepackage{natbib}

\newif\ifAMStwofonts
\AMStwofontstrue

\graphicspath{{./fig/}}

\title[Structure formation in DE models]{A comparison of structure formation in minimally and non-minimally coupled
quintessence models}

\author[Pace et al.]
       {Francesco Pace$^{1}$\thanks{E-mail: Francesco.Pace@port.ac.uk},
	Lauro Moscardini$^{2,3,4}$, Robert Crittenden$^1$, Matthias Bartelmann$^{5}$, \newauthor Valeria
Pettorino$^{6}$\\
	$^{1}$ Institute of Cosmology and Gravitation, University of Portsmouth, Dennis Sciama Building, Portsmouth, PO1
3FX, U.K.\\
	$^{2}$ Dipartimento di fisica e astronomia, Universit\'a di Bologna, Viale Berti Pichat 6/2, 40127 Bologna,
Italy\\
        $^{3}$ INFN, Sezione di Bologna, Viale Berti Pichat 6/2, I-40127 Bologna, Italy\\
        $^{4}$ INAF, Osservatorio Astronomico di Bologna, via Ranzani 1, I-40127 Bologna, Italy\\
	$^{5}$ Zentrum f\"ur Astronomie der Universit\"at Heidelberg, Institut f\"ur Theoretische Astrophysik,
Albert-Ueberle-Str. 2, D-69120 Heidelberg, Germany\\
	$^{6}$ D\'epartement de Physique Th\'eorique and Center for Astroparticle Physics, Universit\'e de Gen\`eve, 24
quai Ernest Ansermet, CH--1211 Gen\`eve 4, Switzerland
}

\date{Received \today; accepted ?}

\pagerange{\pageref{firstpage}--\pageref{lastpage}} \pubyear{2013}

\begin{document}
\label{firstpage}
\maketitle

\begin{abstract}
We study structure formation in non-minimally coupled dark energy models, where there is a coupling in the Lagrangian
between a quintessence scalar field and gravity via the Ricci scalar.
We consider models with a range of different non-minimal coupling strengths and compare these to minimally coupled 
quintessence models with time-dependent dark energy densities. The equations of state of the latter are tuned to
either reproduce the equation of state of the non-minimally coupled models or their background history. Thereby
they provide a reference to study the unique imprints of coupling on structure formation. We show that the coupling
between gravity and the scalar field, which effectively results in a time-varying gravitational constant $G$, is not
negligible and its effect can be distinguished from a minimally coupled model. We extend previous work on this subject
by showing that major differences appear in the determination of the mass function at high masses, where we observe
differences of the order of $40\%$ at $z=0$. Our new results concern effects on the non-linear matter power spectrum and
on the lensing signal (differences of $\approx 10\%$ for both quantities), where we find that non-minimally coupled
models could be distinguished from minimally coupled ones.
\end{abstract}


\section{Introduction}
In recent years, data ranging from observations of Type Ia Supernovae \citep{Riess1998,Perlmutter1999,
Riess2004,Riess2007}, CMB and the integrated Sachs-Wolfe effect
\citep{Jaffe2001,Giannantonio2008,Ho2008,Komatsu2011,Jarosik2011,Planck2013_XV,Planck2013_XVI,Planck2013_XIX}, large
scale structure (LSS) and baryon acoustic oscillations (BAO) \citep{Tegmark2004a,Eisenstein2005,Percival2010}, globular
clusters \citep{Krauss2003}, galaxy clusters \citep{Haiman2001,Allen2004,Allen2008,Wang2004} to weak lensing
\citep{Hoekstra2006,Jarvis2006} and X-ray \citep{Vikhlinin2009} have shown that the expansion rate of the Universe is
presently accelerating. In the framework of General Relativity, this can be explained by supposing that approximately
three quarters of the total energetic budget of the Universe is in the form of an unknown component with negative
pressure, generically known as `dark energy.'

The simplest form of dark energy is the cosmological constant $\Lambda$, a purely geometric term in Einstein's field 
equations, characterized by a constant equation of state ($w=-1$) and so far in agreement with all available
observations. Yet for a cosmological constant ($\Lambda$CDM model), fine-tuning and coincidence problems are quite
severe and remain unsolved.

An alternative is provided by quintessence scalar fields \citep{Wetterich1988,Ratra1988a}. The scalar field is dynamical
and its background evolution is slow enough to closely reproduce the behaviour of the cosmological constant and drive
the accelerated expansion today. However observations constrain quite tightly the equation of state of the dark energy
component to be very close to -1 at present \citep{Komatsu2011} and in this case, as pointed out by \cite{Bludman2004},
the basin of attraction in the early universe is shrinking and thus enhancing the fine tuning present in minimally
quintessence models, as severe as the $\Lambda$CDM one.

Given these considerations, it is worth investigating extensions of General Relativity in which dark energy is
associated to a scalar field non-minimally coupled to gravity. In these extended models, the field dynamics may differ
from that of minimally coupled models due to gravitational effects. In such scenarios, the scalar field mediates a {\it
fifth force}; this happens when there is a universal coupling to all species, as in scalar-tensor theories
\citep{Hwang1991,Demarque1994,Barrow1996,Mashhoon1998,Boisseau2000,Perrotta2000,Faraoni2000,Torres2002,Fujii2003,
Banerjee2009,Jamil2011,Charmousis2012,Wang2012}, or if the coupling is non-universal, as it happens in coupled
quintessence
\citep{Schmidt1990,Wetterich1995,Amendola2000,Holden2000,Sidharth2000,Amendola2003,Matarrese2003,Amendola2004,Mota2008,
Pettorino2008,Guendelman2008,Amendola2008,Zhao2010,Baldi2010,Pettorino2010,Amendola2012,Pettorino2012}; it also occurs 
with physics associated to generalized kinetic energy terms \citep{ArmendarizPicon2001,Caldwell2002,Malquarti2003}.
One effect of this class of models is that the gravitational constant $G$, appearing in Einstein's field equations is 
no longer a constant, but becomes a function of the scalar field and thus becomes time dependent.

In scalar-tensor theories the scalar field is non-minimally coupled to gravity via the Ricci scalar and at present 
times can behave as dark energy. Models with this coupling are also called extended quintessence models
(\cite*{Perrotta2000,Acquaviva2004}, \cite{Acquaviva2005}, \cite*{Pettorino2005b,Pettorino2008}). One of the
consequences of these models is that the coupling of the scalar field to the Ricci scalar in the Lagrangian enhances the
dynamics of the field at early times, an effect known as R-boost \citep*{Baccigalupi2000,Pettorino2005a}. As a
consequence the range of attraction for tracking solutions is conserved also for models where the equation of state is
close to -1 today \citep{Matarrese2004}. Fine tuning is however still present in the choice of a flat potential.

In this paper we investigate the effects of extended quintessence models on structure formation from an analytical 
point of view, thus complementing, validating and extending the work based on N-body numerical simulations by
\cite{DeBoni2011}. The novelty of this work is the study of the spherical collapse in scalar tensor theories
\citep{Bernardeau1994,EspositoFarese2001,Ohta2003,Ohta2004,Mota2004,Perrotta2004,Acquaviva2004,Nunes2006,Abramo2007,
Pettorino2008,Basilakos2009, Pace2010, Basilakos2010, Wintergerst2010a}. To this purpose we generalize the
semi-analytical spherical collapse model to take into account effects from the scalar field (which, for simplicity is
considered to be homogeneous) in order to study the time behaviour of the linearly extrapolated density contrast
$\delta_{\mathrm{c}}$ and the linear growth factor. We will study five minimally coupled dark energy models, two of
which with the same equation-of-state parameter of the simulated extended quintessence models and two with an equation
of state tuned to reproduce the same background history of the simulated non-minimally coupled models.

The paper is organized as follows: 
In Section~\ref{sect:models} we present the models studied in this work and we describe how to take into account the 
scalar field for perturbation theory in the quasi-static Newtonian regime in scalar-tensor theories. In
Section~\ref{sect:results} we present our results for the linear growth
factor (Section~\ref{sect:gf}), the spherical collapse parameters $\delta_{\rm c}$ and $\Delta_{\rm V}$
(Section~\ref{sect:spcstt}), the mass function (Section~\ref{sect:mf}), the non-linear matter power spectrum
(Section~\ref{sect:DMps}) and the cosmic shear power spectrum (Section~\ref{sect:lensing}). Finally
Section~\ref{sect:conc} is devoted to our conclusions.
Throughout we work in units where the speed of light is $c=1$.

\section{Cosmological models}\label{sect:models}
\subsection{$\Lambda$CDM and quintessence dark energy models}
In this work we will consider as fiducial model the $\Lambda$CDM model, characterized by the presence of a 
cosmological constant described by an equation of state $w=-1$, constant at all times. This implies that the amount of 
dark energy will not change and eventually comes to dominate the total energy density. In other dark energy models we
consider, the equation-of-state parameter is in general a function of time. In a homogeneous and isotropic Universe, the
cosmological expansion can be written in terms of the first Friedmann equation
\begin{equation}
H^{2}=H_{0}^{2}\left[\Omega_{\mathrm{r},0}a^{-4}+\Omega_{\mathrm{m},0}a^{-3}+\Omega_{\mathrm{K},0}a^{-2}+
\Omega_{\mathrm{q},0}g_{\mathrm{q}}(a)\right]\;,
\label{eqn:H}
\end{equation}
where $\Omega_{\mathrm{r},0}$ represents the radiation, $\Omega_{\mathrm{m},0}$ the matter, $\Omega_{\mathrm{K},0}$ the 
curvature and $\Omega_{\mathrm{q},0}$ the dark energy densities today, respectively. The function $g(a)$ describes the 
time evolution of the dark energy density component. For a perfect fluid, where the pressure ($P$) and energy density
($\rho$) are related by some dark energy equation of state, $P=w(a)\rho$, $g_{\mathrm{q}}(a)$ is
\begin{equation}
g_{\mathrm{q}}(a)=\exp{\left(-3\int_{1}^{a}\frac{1+w(a^{\prime})}{a^{\prime}}da^{\prime}\right)}\;.\label{eqn:g}
\end{equation}
As one can easily see from Eq.~\ref{eqn:g}, for the cosmological constant $g_{\mathrm{q}}(a)=1$.

The idea of replacing the cosmological constant by the energy density of a scalar field was explored in several works
\citep[][]{Wetterich1985,Wetterich1988,Wetterich1995,Ratra1988a} and if the scalar field does not experience any 
direct coupling to any of the other constituents of the models it is said to be minimally coupled and the action reads
\begin{equation}\label{eqn:mca}
S=\int d^{4}x\sqrt{-g}\left(\frac{R}{16\pi G}+\mathcal{L}_{\phi}+\mathcal{L}_{\rm fl}\right)\;,
\end{equation}
where $g$ is the determinant of the metric, $R$ the Ricci scalar, $\mathcal{L}_{\rm fl}$ is the Lagrangian of all
fluids except the dark energy scalar field and $\mathcal{L}_{\phi}$ represents the Lagrangian of the scalar field
\begin{equation}
\mathcal{L}_{\phi}=-\frac{1}{2}\nabla^{\mu}\phi\nabla_{\mu}\phi-V(\phi)\;,
\end{equation}
where $V(\phi)$ denotes the self-interaction potential of $\phi$. Below we will generally assume the potential takes 
the \emph{Ratra-Peebles} form, 
\begin{equation}
V(\phi)=\frac{M^{4+\alpha}}{\phi^{\alpha}}\;,
\end{equation}
where $M$ is a typical energy scale and $\alpha$ is a free positive exponent.

Varying the action in Eq.~\ref{eqn:mca} with respect to the metric $g_{\mu\nu}$ gives the usual Einstein field 
equations
\begin{equation}
G_{\mu\nu}=8\pi G\left[T_{\mu\nu}^{(\mathrm{fl})}+T_{\mu\nu}^{(\phi)}\right]\;,
\end{equation}
where $G_{\mu\nu}$ is the Einstein tensor, $T_{\mu\nu}^{(\mathrm{fl})}$ is the stress-energy tensor for a 
homogeneous and isotropic cosmic fluid (here dominated by dark matter) and $T_{\mu\nu}^{(\phi)}$ is the stress-energy 
tensor for the quintessence scalar field:
\begin{equation}\label{eqn:Tmn}
T_{\mu\nu}^{(\phi)}=\nabla_{\mu}\phi\nabla_{\nu}\phi-g_{\mu\nu}\left(\frac{1}{2}\nabla^{\alpha}\phi\nabla_{\alpha}
\phi+V(\phi)\right)\;.
\end{equation}

Assuming a spatially flat Friedmann-Robertson-Walker (FRW) metric $ds^{2}=-dt^{2}+a^{2}(t)d\vec{x}^{2}$ where $a(t)$ is 
the scale factor we can identify the energy density and pressure of the scalar field as
\begin{eqnarray}
\rho_{\phi} & = & \frac{1}{2}\dot{\phi}^{2}+V(\phi)\label{eqn:rhophi} \\ 
p_{\phi} & = & \frac{1}{2}\dot{\phi}^{2}-V(\phi)\label{eqn:Pphi} \;.
\end{eqnarray}
Varying the action $S$ with respect to the scalar field itself we derive the equations of motion which resemble the
Klein-Gordon equation for a spatially homogeneous field on an isotropically expanding space-time
\begin{equation}
\ddot{\phi}+3H\dot{\phi}+\frac{dV(\phi)}{d\phi}=0\;.
\end{equation}
With the assumption of a flat FRW metric, the scalar field satisfies the continuity equation
\begin{equation}
\dot{\rho}_{\phi}+3H(\rho_{\phi}+p_{\phi})=0\;,
\end{equation}
so that we can write $\rho_{\phi}=\rho_{\phi,0}g_{\mathrm{q}}(a)$, with $g_{\mathrm{q}}(a)$ defined in Eq.~\ref{eqn:g}.

\subsection{Scalar-tensor models}
Scalar-tensor (sometimes called extended quintessence) models are instead described by the action
\begin{equation}\label{eqn:nmca}
S=\int d^{4}x\sqrt{-g}\left(\frac{1}{2}f(\phi,R)+\mathcal{L}_{\phi}+\mathcal{L}_{\mathrm{fl}}\right)\;,
\end{equation}
where this formulation was first introduced in a cosmological context by \cite{Hwang1991}. 
With respect to General Relativity, the term $R/16\pi G$ is replaced by an arbitrary function of the Ricci scalar and 
scalar field $f(\phi,R)/2$. 
In addition, the scalar field is described by the Lagrangian
\begin{equation}
\mathcal{L}_{\phi} =-\frac{1}{2}\omega(\phi)\nabla^{\mu}\phi\nabla_{\mu}\phi-V(\phi)\;,
\end{equation}
where $\omega(\phi)$ is a function of the scalar field only which generalizes the kinetic term.

These models are interesting because they are related to the original Brans-Dicke idea \citep{Brans1961} and to the 
attempt to explain cosmic acceleration exclusively in terms of modifications of General Relativity.
Such models have been studied in several works \cite[see
also][]{Wetterich1995,Barrow1997,Sahni1998,Uzan1999,Bartolo2000,Boisseau2000,
Perrotta2000,Faraoni2000,EspositoFarese2001,Torres2002,Perrotta2002,Linder2004,Matarrese2004,Pettorino2005a,
Pettorino2008,Tsujikawa2008,Boisseau2011,BuenoSanchez2011,Jamil2011,Charmousis2012,Wang2012}.
Here we just summarize the most important aspects that will be relevant for the present work.

The variation of the action described in Eq.~\ref{eqn:nmca} with respect to the metric $g_{\mu\nu}$ yields the field
equations
\begin{equation}\label{eqn:EFE}
G_{\mu\nu}=8\pi
GT_{\mu\nu}=\frac{1}{f^{\prime}}\left[T_{\mu\nu}^{(fl)}+T_{\mu\nu}^{(\phi)}+\frac{1}{2}g_{\mu\nu}(f-f^{\prime}R)+
A_{\mu\nu}(f^{\prime})\right]\;
\end{equation}
where the tensor $A_{\mu\nu}$ is defined for an arbitrary scalar $h$ as
\begin{equation}
A_{\mu\nu}(h)=\nabla_{\mu}\nabla_{\nu}h-g_{\mu\nu}\Box h\;,
\end{equation}
and $f^{\prime}\equiv\partial f/\partial R$.
The Klein-Gordon equation for the scalar field is also modified with respect to the minimally coupled case,
\begin{equation}
\ddot{\phi}+3H\dot{\phi}=-\frac{1}{2\omega}\left(\frac{d\omega}{d\phi}\dot{\phi}^2-
\frac{\partial f}{\partial\phi}R+2\frac{dV}{d\phi}\right)\;.
\end{equation}
The energy-momentum tensor for the scalar field now reads
\begin{equation}
T_{\mu\nu}^{\phi}=\omega(\phi)\left(\nabla_{\mu}\phi\nabla_{\nu}\phi-
\frac{1}{2}g_{\mu\nu}\nabla^{\alpha}\phi\nabla_{\alpha}\phi\right)-g_{\mu\nu}V(\phi)\;,
\end{equation}
and it is interesting to note that, unlike in the minimally coupled models, these modifications imply that the energy 
density of the scalar field no longer satisfies the continuity equation for the background quantities
$(\dot{\rho}+3H(\rho+p)=0)$ \citep{Hwang1991}.

We assume that $f(\phi,R)$ is linear in the Ricci scalar, 
\begin{equation} 
f(\phi,R)=\frac{F(\phi)}{8\pi G_{\ast}}R; 
\end{equation} 
then by identifying the energy-momentum tensor of the scalar field with that of a perfect fluid, we can derive
expressions for the non-conserved background energy density and pressure of the scalar field:
\begin{eqnarray}
\rho_{\phi} & = & \frac{1}{2}\omega(\phi)\dot{\phi}^{2}+V(\phi)-3H\dot{F}(\phi)\\
p_{\phi} & = & \frac{1}{2}\omega(\phi)\dot{\phi}^{2}-V(\phi)+\ddot{F}(\phi)+2H\dot{F}(\phi)\;.
\end{eqnarray}
These models can be related to the original Brans-Dicke gravity \citep{Brans1961} when $F=\phi$ and
$\omega(\phi)=\omega_{\mathrm{BD}}\phi$, while non-minimally coupled theories are obtained with the identifications
$\omega(\phi)= F(\phi) = 1$.

With this linear ansatz for $f(\phi,R)$, Friedmann's equations become
\begin{eqnarray}
H^{2} & = & \frac{8\pi G}{3F}\left(\rho_{\mathrm{fl}}+\frac{1}{2}\dot{\phi}+V(\phi)-3H\dot{F}\right)\\
\frac{\ddot{a}}{a} & = & -\frac{4\pi G}{3F}\left[\rho_{\mathrm{fl}}+3p_{\mathrm{fl}}+2\dot{\phi}^{2}-2V(\phi)-3(\ddot{F}
+H\dot{F})\right]\;.
\end{eqnarray}
If we define an effective Jordan-Brans-Dicke parameter as, 
\begin{equation}\label{eqn:chi-omega}
\omega_{JBD}=\frac{F(\phi)}{(F_{,\phi}(\phi))^2}\;,
\end{equation}
then General Relativity is recovered when $\omega_{JBD}\gg 1$.

Any changes to the gravitational physics require matching the GR behaviour on solar system scales where gravity is well 
tested, while still reproducing the observed effects of dark energy on large scales. Here we require that
\begin{equation}
f(\phi_0,R)=\frac{R}{8\pi G}\;,
\end{equation}
where $\phi_{0}$ is the value of the scalar field today and $G$ is the gravitational constant measured today.
In the following we will assume the modifications take the following form:
\begin{itemize}
\item[-] $\omega(\phi)=1$
\item[-] $F(\phi)=1+8\pi G_{\ast}\xi(\phi^{2}-\phi_{0}^{2})$
\end{itemize}
where $\xi$ is the coupling constant, and $\phi_0$ is the present value of the scalar field. 
$G_{\ast}$ is the bare gravitational constant \citep{EspositoFarese2001} which
in general differs from the gravitational constant $G$ appearing in Einstein's or Newton's field equations.

Very tight constraints ($\xi\approx 10^{-2}$) on the coupling parameter come from solar system tests
\citep{Reasenberg1979,Uzan1999,Chiba1999,Will2001,Riazuelo2002,Bertotti2003}, from cosmological scale measurements
\citep{Clifton2005,Acquaviva2005,Appleby2010,Farajollahi2011} and nucleosynthesis
\citep{Accetta1990,Torres1995,Santiago1997,Coc2006,Lee2011}.
These works assume no screening mechanism, while other works assume screening, either exploiting the chameleon effect 
\citep{Mota2004b,Khoury2004} or the Vainshtein mechanism \citep{Vainshtein1972}. Many simulations now take into account
such screening mechanism \citep{Oyaizu2008,Schmidt2009,Zhao2011}.

\subsection{Model parameters}\label{sect:models_params}
We will compare analytic results presented in Section~\ref{sect:mf} with N-body simulations discussed in
\cite{DeBoni2011}. To do so, we adopt the WMAP \citep{Spergel2007} cosmological parameters which had been used in these
simulations ($\Omega_{\mathrm{m,0}}=0.268$, $\Omega_{\phi,0}=0.732$, $h_{0}=0.704$) and throughout the paper we assume a
spatially flat cosmological background, $\Omega_{\mathrm{K,0}}=0$. 
These parameters are slightly different from the ones found by Planck\footnote{planck.esa.int/}, but here we want to
emphasise the comparison of models having the same cosmological parameters, therefore this does not represent an issue
for the conclusions of our work.

We consider two non-minimally coupled models and five minimally coupled models (see Table~\ref{tab:params}).
These models are labelled  NMC$n$, MCw$n$ or MCH$n$ respectively, where the index $n$ runs from one to two.
The non-minimally coupled models differ essentially in their coupling strengths. They correspond to the maximum
deviation from General Relativity allowed by current observations on cosmological scales. The minimally coupled models
MCw1 and MCw2 have the same equation of state as the simulated extended quintessence models (NMC1 and NMC2) and the
minimally coupled models MCH1 and MCH2 have the same background history of the models NMC1 and NMC2 in order to
independently evaluate the effect of the coupling and of the time dependent gravitational constant. A fifth model, wCDM,
has a constant equation of state parameter $w=-0.9$, the highest value consistent with observational constraints
\citep{Unnikrishnan2008}.

We normalize the amplitude of the primordial power spectrum for the fiducial $\Lambda$CDM cosmology to have a value of 
the quadratic deviation on a comoving scale of 8~Mpc$/h$ of $\sigma_{8}=0.776$. Dark energy models are normalized 
to match the amplitude of fluctuations at the CMB epoch $z_{\mathrm{CMB}}=1089$ according to the relation
\begin{equation}\label{eqn:sigma8}
\sigma_{8,\mathrm{DE}}=\sigma_{8,\Lambda\mathrm{CDM}}\frac{D_{+,\Lambda\mathrm{CDM}}(z_{\mathrm{CMB}})}{D_{+,\mathrm{DE}
}(z_{\mathrm{CMB}})}\;.
\end{equation}
In the previous equation $D_{+}(z)$ is the linear growth factor normalised to unity today (see Sect.~\ref{sect:gf}). 
An alternative normalization which is often adopted in the literature is to fix the exponential tail of the mass 
function to be approximately the same at $z=0$; therefore differences will arise at earlier times. 
However, since the normalization of the fluctuations is bounded to high accuracy by the CMB measurements, we 
exclusively adopt the first normalization. The values of the parameters for the extended quintessence models are chosen
so that the energy density of the scalar field today is approximately the same as that of the cosmological constant. The
other differences arising at $z=0$ may be used to discriminate among the different cosmological models.

\begin{table}
\caption{Values of the parameters adopted for the reference $\Lambda$CDM model and the dynamical dark energy models. 
The exponent of the inverse power-law potential is indicated with $\alpha$; $\xi$ is the strength of the coupling, 
$\omega_{\mathrm{JBD},0}$ is the present value of the effective {Jordan-Brans-Dicke parameter} and $\sigma_{8}$ 
represents the normalization of the matter power spectrum such that fluctuations are the same at $z_{\mathrm{CMB}}$.}
\label{tab:params}
\begin{center}
\begin{tabular}{c|c|c|c|c|c}
\hline
\hline
Model & $\alpha$ & $\xi$ & $\omega_{\mathrm{JBD},0}$ & $\sigma_{8}$\\
\hline
$\Lambda$CDM & - & - & - & 0.776 \\
NMC1 & 0.229 & +0.085 & 120 & 0.748 \\
NMC2 & 0.435 & -0.072 & 120 & 0.729 \\
MCw1 & - & - & - & 0.752 \\
MCw2 & - & - & - & 0.745 \\
MCH1 & - & - & - & 0.744 \\
MCH2 & - & - & - & 0.760 \\
wCDM & - & - & - & 0.753 \\
\hline
\hline
\end{tabular}
\end{center}
\end{table}

\begin{figure}
\includegraphics[angle=-90,width=\hsize]{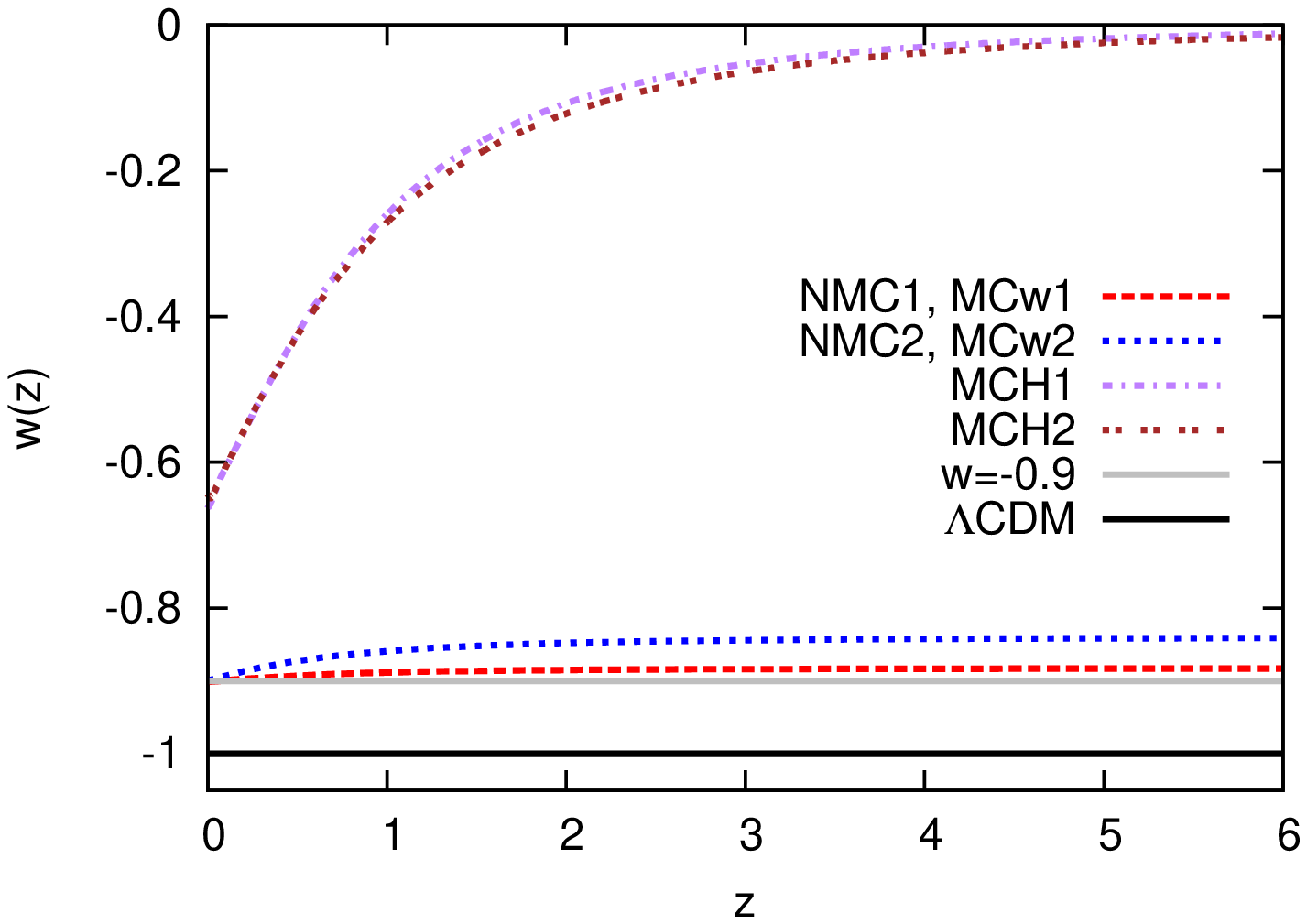}
\includegraphics[angle=-90,width=\hsize]{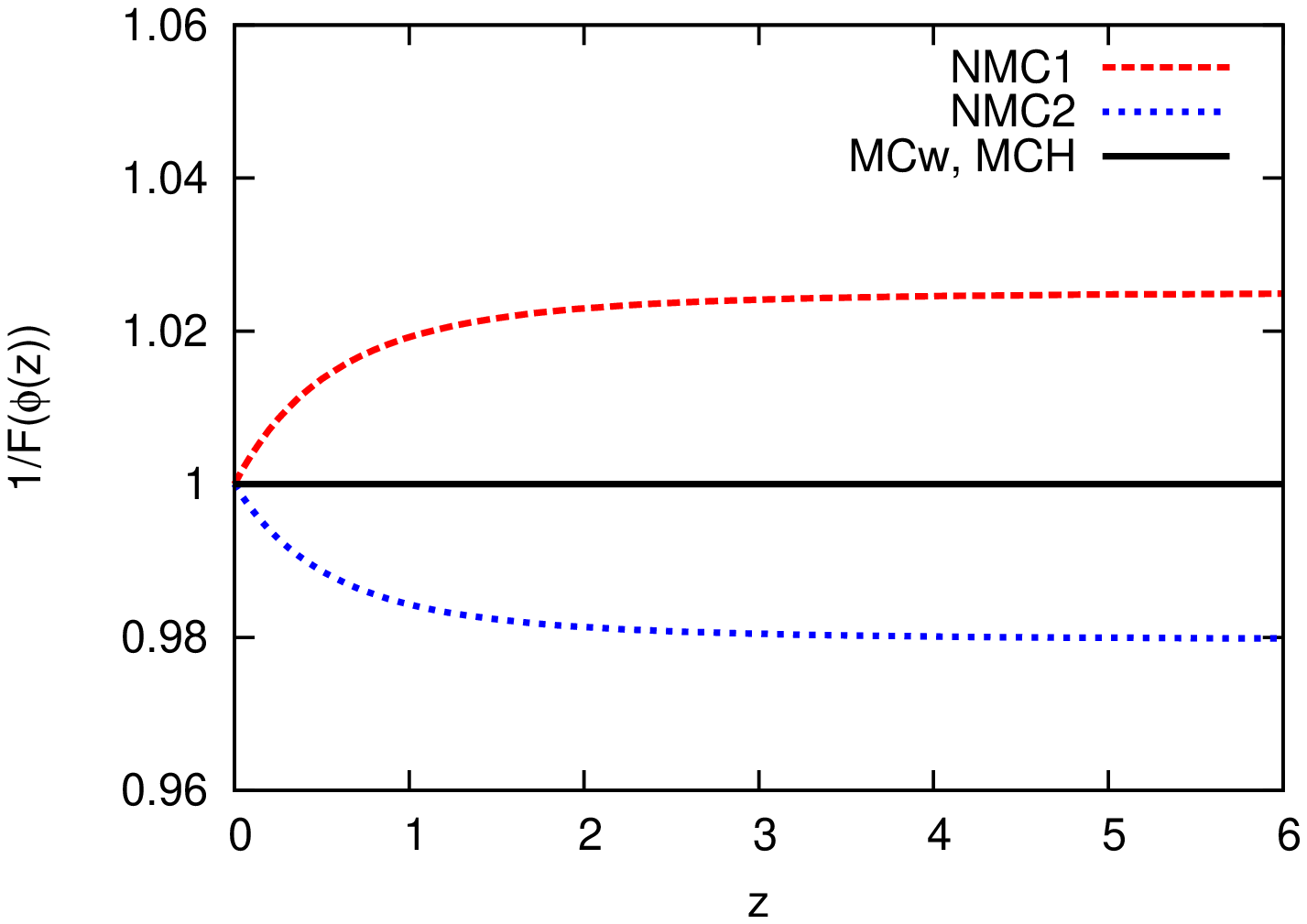}
\caption{Upper panel: equation of state for the dark energy models considered in this work as a function of redshift. 
Lower panel: redshift dependence of the function $F^{-1}(\phi) \simeq 1-8\pi G_\ast \xi(\phi^2-\phi_0^2)$. The red 
dashed curve represents the model NMC1 with coupling constant $\xi=0.085$, the blue short-dashed curve represents the
model NMC2 with coupling constant $\xi=-0.072$. The two minimally coupled models MCw1 and MCw2 have the same
equation-of-state parameter of the extended quintessence models NMC1 and NMC2 and are shown with the same curve. Models
MCH1 and MCH2 are shown with violet short-dashed-dotted and brown dot-dotted line. Finally, the reference $\Lambda$CDM
($w=-1$) and wCDM ($w=-0.9$) models are shown with a black and grey solid line, respectively.}
\label{fig:wF}
\end{figure}

In Fig.~\ref{fig:wF} we show the redshift evolution of the equation of state $w$ (upper panel) and of the function
$1/F(\phi)$ (lower panel) for the quintessence models studied in this work. We refer to the caption for the different
colours and line styles adopted. The value of the equation of state at $z=0$ is close to $w=-0.9$ for all the dynamical
models investigated, except for the models MCH1 and MCH2. The equations of state become essentially constant for $z>3$.
The minimally coupled dark energy models MCw1 and MCw2 are described by the same $w$ as models NMC1 ($\xi=0.085$) and
NMC2 ($\xi=-0.072$).
The equation of state for the minimally coupled models MCH$n$ is derived using the following expression:
\begin{equation}
 w(a)=-\frac{1+\frac{2}{3}a\frac{d\ln E(a)}{da}+\frac{1}{3}\frac{\Omega_{\rm{r}}}{a^4E(a)^2}-
 \frac{1}{3}\frac{\Omega_{\rm{K},0}}{a^2E(a)^2}}{1-\frac{\Omega_{\rm{m},0}}{a^3E(a)^2}-
 \frac{\Omega_{\rm{r}}}{a^4E(a)^2}-\frac{\Omega_{\rm{K},0}}{a^2E(a)^2}}\;,
\end{equation}
where $H(a)=H_0E(a)$.

The function $F(\phi)$ changes rapidly at low redshifts and becomes practically constant for $z\gtrsim 2$, differing 
from the minimally coupled case by at most $2.5\%$. Since in the field equations the usual gravitational 
constant $G$ is replaced by the function $1/F(\phi)$, according to the sign of the coupling constant, gravity will be 
stronger ($\xi>0$) or weaker compared ($\xi<0$) to the minimally coupled case. This happens because of the functional
form of $F(\phi)=1+8\pi G_\ast\xi(\phi^{2}-\phi_0^2)$.

\subsection{Background properties}
\begin{figure}
\includegraphics[angle=-90,width=0.95\hsize]{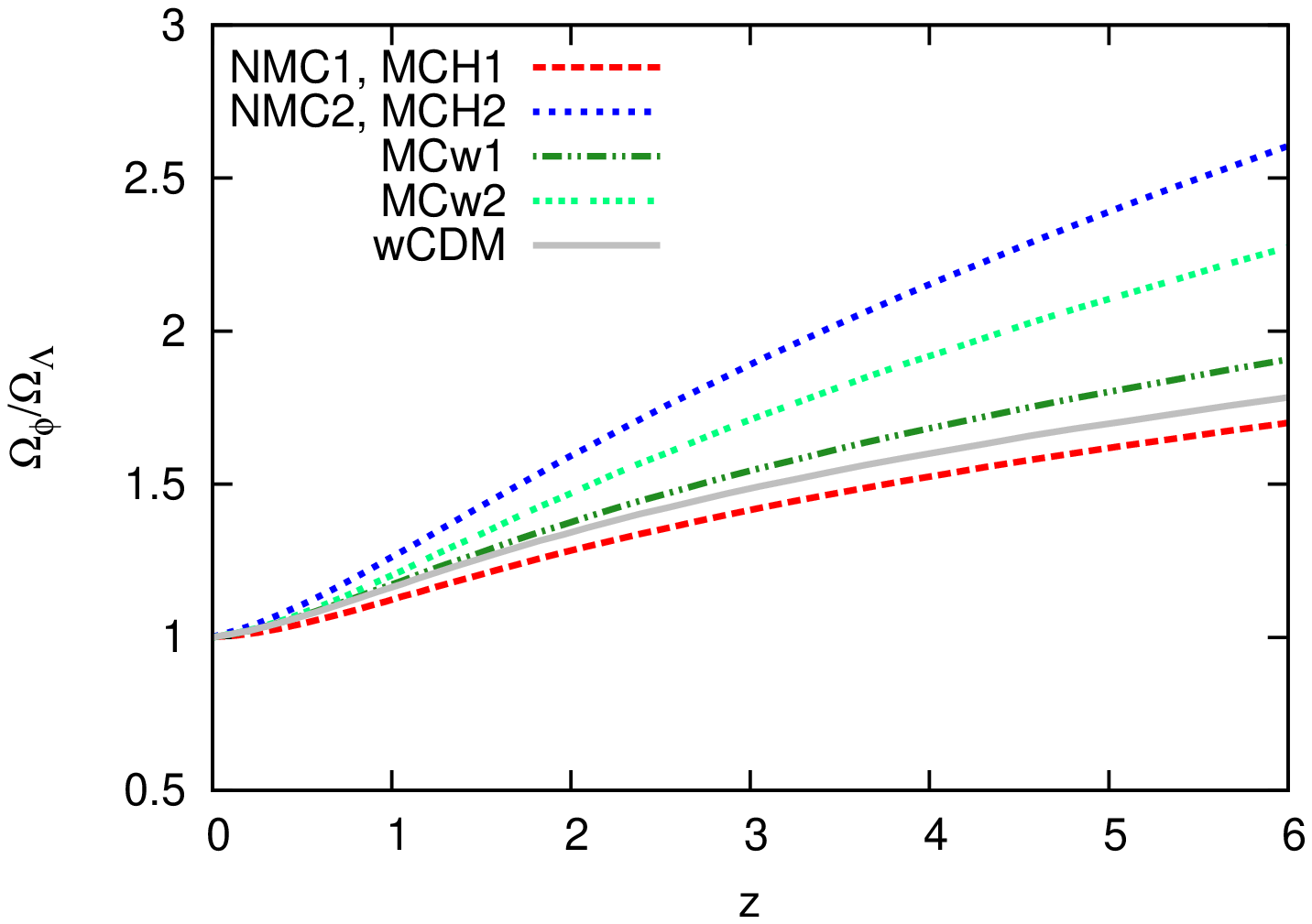}
\includegraphics[angle=-90,width=0.95\hsize]{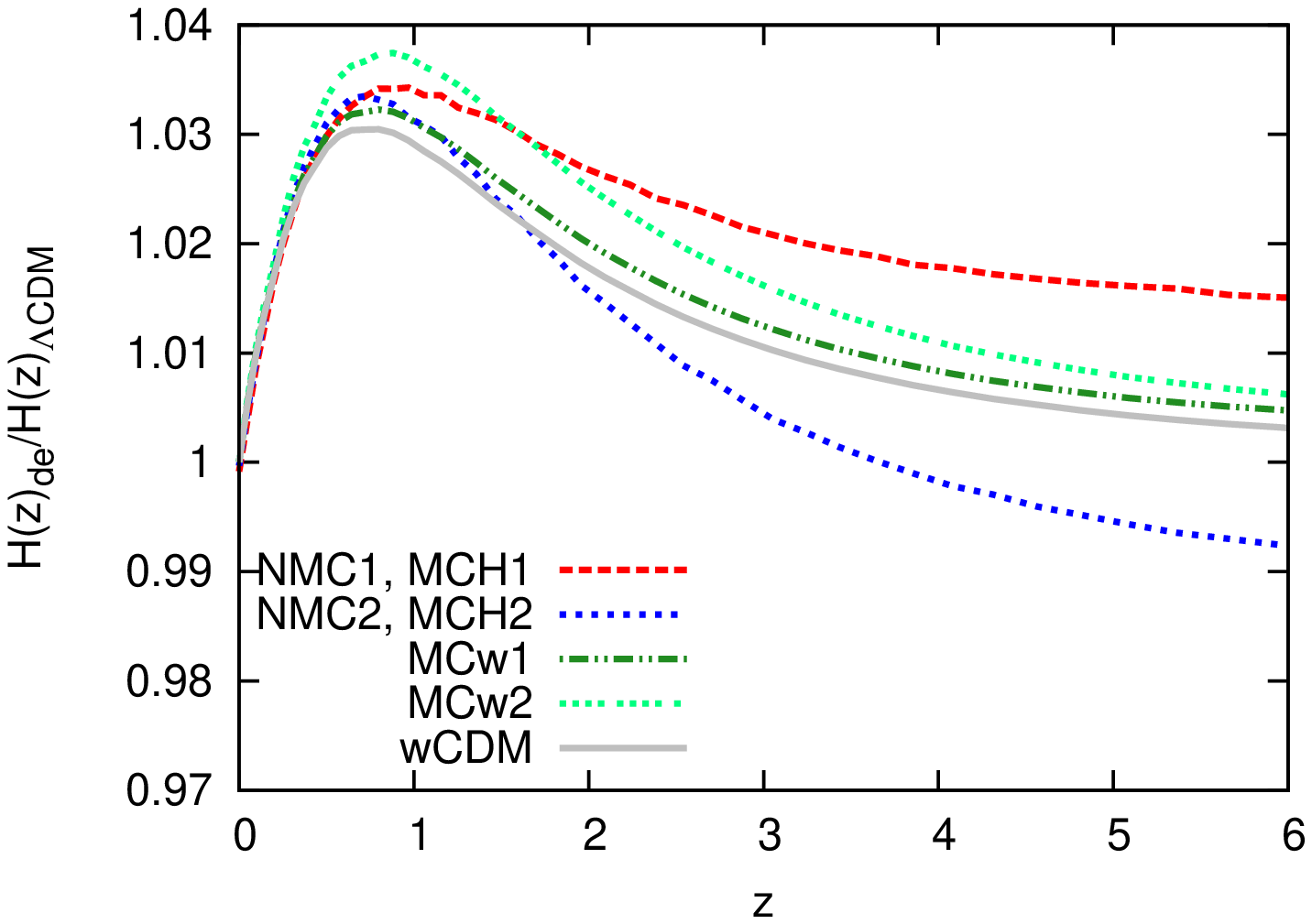}
\caption{Upper (lower) panel: redshift evolution of the ratio of the scalar field (Hubble) parameter for the 
different models studied to the corresponding quantity in the fiducial $\Lambda$CDM model. Line types and colours for
the non-minimally coupled models are as in the upper panel of Fig.~\ref{fig:wF}. Models MCw1 and MCw2 are shown with
dark dot-dashed-dotted and light green dotted curves, respectively.}
\label{fig:bck}
\end{figure}

As the values of the coupling constants are small, we expect small differences at the background level between these 
models and the reference $\Lambda$CDM model. Our expectations are confirmed by Fig.~\ref{fig:bck} where we show the 
ratio of the scalar field density and the Hubble parameter (upper and lower panel respectively) as a function of
redshift for the eight models considered here with respect to the cosmological constant model. For the Hubble parameter,
the maximum difference, $\approx 4\%$, takes place at $z\approx 1$. These differences in the Hubble function render
the differences in the age of the Universe or in distances to be very small, of the order of a few percent. Similar
differences are present in the comoving and in the luminosity distance. Ratios of the matter density fraction 
will be simply related to the corresponding Hubble functions.

Since the Hubble parameter is a key ingredient in determining the time evolution of the perturbations, we can infer that
they will not substantially differ from results expected in the $\Lambda$CDM case, as explained in detail in the
following sections.

It is worth noticing that the minimally coupled models MCw1, MCw2 and wCDM have very similar expansion histories, while
the other models differ more. This is due to the fact that they are characterised by the same coupling constant, and
arises despite the fact that the dark energy equations of state are explicitly tuned to match. This shows the
importance, already at the background level, of the coupling constant. In fact we can also see from the lower panel of
Fig.~\ref{fig:bck} that models having the same coupling constants show very similar expansion histories and time
evolutions of the matter content.

\subsection{Perturbations}\label{sect:pstt}
We now review the main features of linear perturbation theory within non-minimally coupled cosmologies in the 
Newtonian limit. For an extended review we refer the reader to \cite{Pettorino2008} and \cite{Wintergerst2010a}.

In the Newtonian limit, time derivatives are negligible with respect to spatial derivatives and the condition $k\gg H$ 
holds. In other words we are considering a quasi-static regime and that scales of interest are much smaller than the
horizon. In this limit, the perturbed continuity, Euler and Poisson equations are (in comoving coordinates)
\begin{eqnarray}
\dot{\delta} & = & -\vec{\nabla}_{\vec{x}}\cdot\vec{u}\\
\frac{\partial\vec{u}}{\partial t} & = & -2H\vec{u}-\frac{1}{a^2}\nabla_{\vec{x}}\psi\\
\vec{\nabla}_{\vec{x}}^{2}\phi_{\mathrm{E}} & = & \frac{4\pi G}{F}\bar{\rho}_{\mathrm{m}}\Omega_{\mathrm{m}}\delta\;,
\end{eqnarray}
where $\delta$ is the matter perturbation, $\vec{u}$ the comoving peculiar velocity and $\psi$ is the Newtonian 
gravitational potential. The potential $\phi_{\mathrm{E}}$ appearing in the Poisson equation is defined as
\begin{equation}\label{eqn:phiE}
\phi_{\mathrm{E}}=\left(1+\frac{1}{2}\frac{F_{,\phi}^2}{F+F_{,\phi}^2}\right)\phi\;.
\end{equation}
The Poisson equation can be rewritten also in terms of the gravitational potential $\psi_{\rm E}$
\begin{equation}\label{eqn:PphiE}
 \vec{\nabla}_{\vec{x}}^{2}\psi_{\rm E}=-\frac{4\pi G}{F}\bar{\rho}_{\mathrm{m}}\Omega_{\mathrm{m}}\delta\;,
\end{equation}
where $\psi_{\rm E}$ is defined as
\begin{equation}\label{eqn:psiE}
 \psi_{\rm E}=\left(1-\frac{1}{2}\frac{F_{,\phi}^2}{F+2F_{,\phi}^2}\right)\psi\;.
\end{equation}
The Euler equation can therefore be modified to
\begin{equation}
 \frac{\partial\vec{u}}{\partial t}+2H\vec{u}+
\left(1+\frac{F_{,\phi}^2}{F+F_{,\phi}^2}\right)\vec{\nabla}_{\vec{x}}\phi=0\;.
\end{equation}
Combining now all the three equations, we obtain a second order differential equation describing the time evolution of 
the linear growth factor:
\begin{equation}\label{eqn:gf}
\ddot{\delta}+2H\dot{\delta}-4\pi G_{\mathrm{eff}}\bar{\rho}_{\mathrm{m}}\delta=0\;,
\end{equation}
where $G_{\mathrm{eff}}$ is defined as (see also \cite{EspositoFarese2001})
\begin{equation}\label{eqn:Geff}
G_{\mathrm{eff}}=\frac{G}{F}\frac{2(F+F_{,\phi}^2)}{2F+3F_{,\phi}^2}\approx\frac{G}{F}\;,
\end{equation}
for $\xi\ll 1$.

As the coupling constant is $\xi\ll 1$, in our models, we can use the approximation $G_{\rm eff}=G/F$. The equation for 
the growth factor is similar the one obtained in $f(R)$ models. This approximation is valid since, in our models,
$F_{,\phi}/F\ll 1$.

\section{Results}\label{sect:results}
In this section we will present results concerning structure formation for the quintessence models described 
above in both the linear and non-linear regimes. In particular we study the growth factor (Sect.~\ref{sect:gf}), the
linear and non-linear overdensity parameter (Sect.~\ref{sect:spcstt}), the mass function (Sect.~\ref{sect:mf}), the
analytical non-linear dark matter power spectrum (Sect.~\ref{sect:DMps}) and the cosmic shear power spectrum
(Sect.~\ref{sect:lensing}).

\subsection{Growth factor}\label{sect:gf}
\begin{figure}
\includegraphics[angle=-90,width=0.95\hsize]{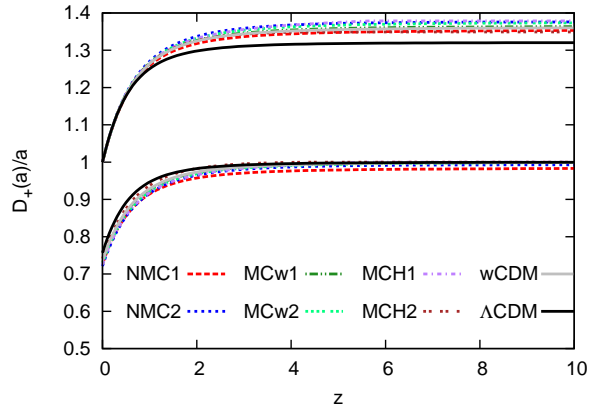}
\caption{The linear growth factor divided by the scale factor $D_{+}(a)/a$ as a function of the redshift. Upper (lower) 
curves show the linear growth factor normalised to unity today (at the CMB epoch). Line styles and colours for the
quintessence models are the same as in Fig.~\ref{fig:bck}, while the fiducial $\Lambda$CDM model is shown with the solid
black line.}
\label{fig:gf}
\end{figure}

The linear growth factor has been extensively studied in several works \citep[see
e.g.,][]{Copeland2006,Perivolaropoulos2007,Tsujikawa2008,Pettorino2008,Lee2011,BuenoSanchez2011}. In Fig.~\ref{fig:gf} 
we show the growth factor divided by the scale factor ($D_{+}(a)/a$) for the dark energy models considered in this work,
as compared to the fiducial $\Lambda$CDM model. We show two normalisations for the fluctuations, matching the amplitudes
at the present time ($z=0$) or at the last scattering of the CMB (see e.g.~\cite{Bartelmann2006}).

\begin{figure}
\includegraphics[angle=-90,width=0.95\hsize]{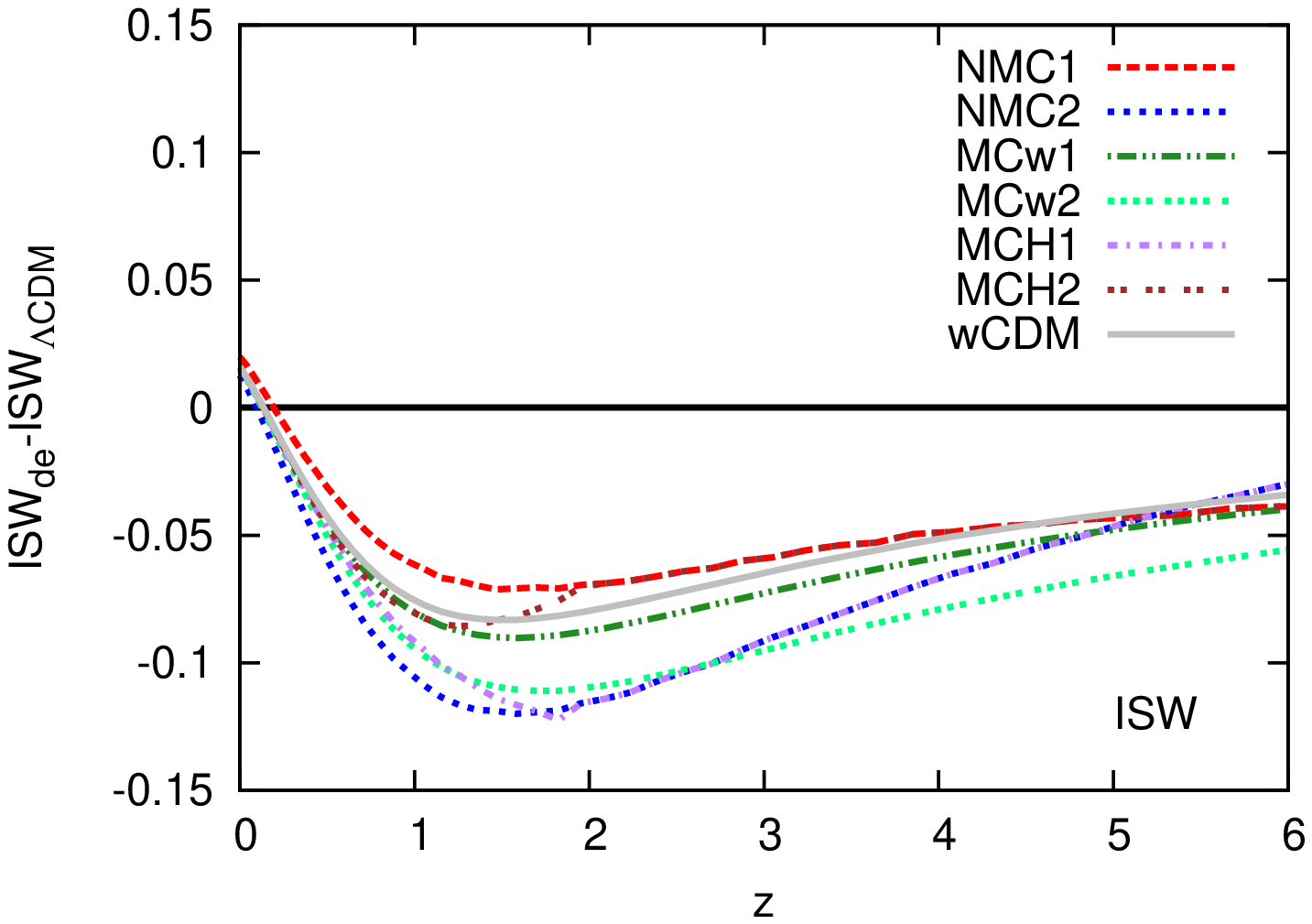}
\includegraphics[angle=-90,width=0.95\hsize]{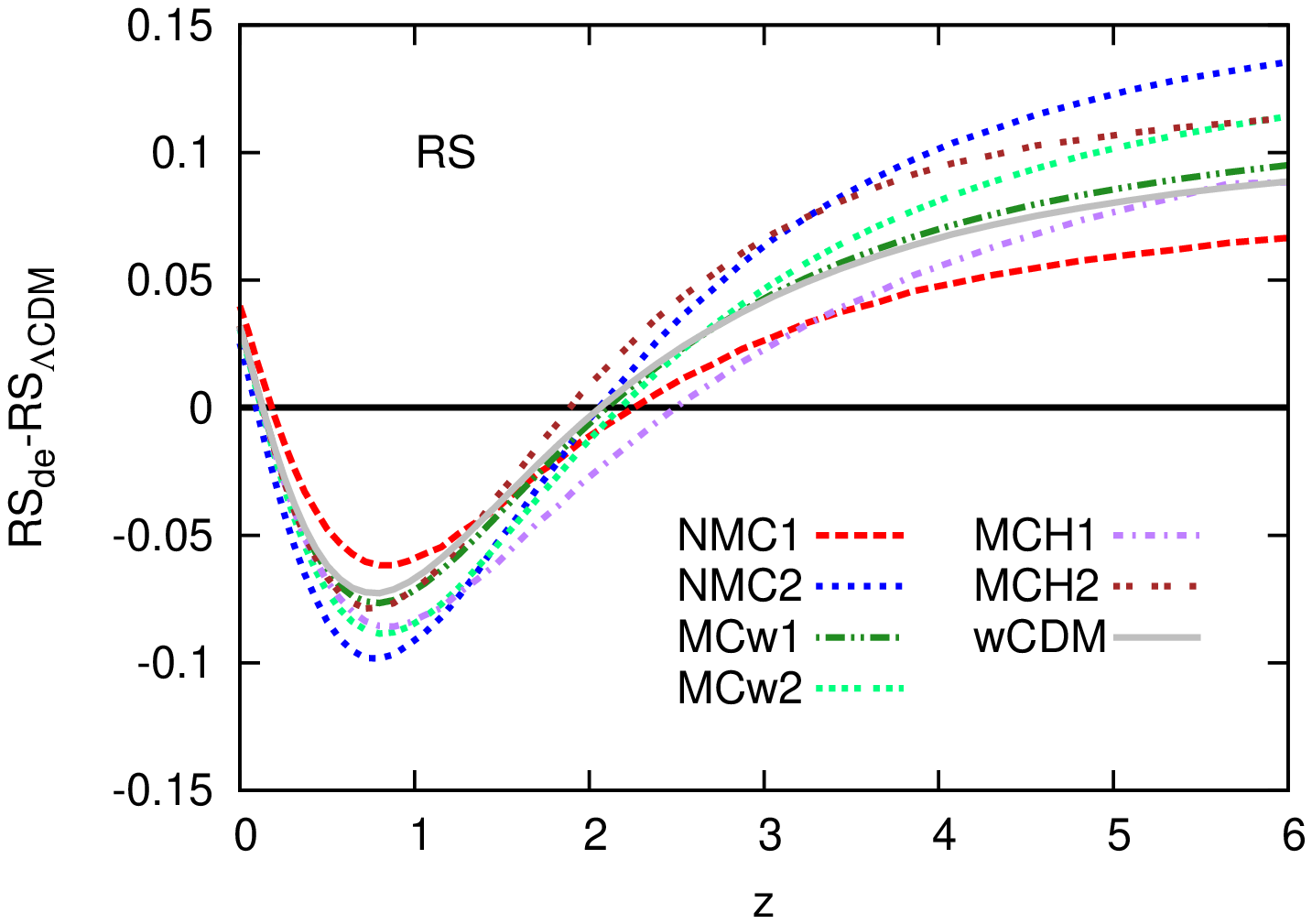}
\caption{Upper (lower) panel: redshift evolution of the difference for the quantity characterising the ISW
(Rees-Sciama) signal between the dark energy models and the $\Lambda$CDM one. Line styles and colours for
the quintessence models are the same as in Fig.~\ref{fig:gf}.}
\label{fig:isw_rs}
\end{figure}

We observe that differences between the dark energy models and the $\Lambda$CDM model are of few percent, ranging
between $2\%$ for the NMC1 model and $5\%$ for the NMC2 model. The minimally coupled model MCH1
(MCH2) behaves very similarly to the non-minimally coupled model NMC2 (NMC1). This is easily explained taking into
account that the source term in Eq.~(\ref{eqn:gf}) is modified by the coupling function $F(\phi)$ and this function
compensates the differences in the background expansion history.
Generally, the quintessence models show less growth compared to the $\Lambda$CDM model. 
When the growth factor is normalised to unity now, primordial perturbations have to be higher to give the same number of
structures today. When instead the growth factor is normalised to unity at early times, the growth factor is lower
because the higher amount of dark energy slows down structure growth.

The study of the growth factor is relevant also for the evaluation of the integrated Sachs-Wolfe (ISW) 
\citep{Sachs1967} and of the Rees-Sciama (RS) effects \citep{Rees1968}.
The ISW effect is due to the interaction of CMB photons with a time varying gravitational potential. The relative 
change of the CMB temperature is given by
\begin{equation}
 \tau=\frac{\Delta T}{T_{\rm CMB}}=\frac{2}{c^3}\int_{0}^{\chi_{\rm H}}d\chi a^2H(a)
 \frac{\partial}{\partial a}(\Phi-\Psi)\;,
\end{equation}
where $\chi_{\rm H}$ is the horizon distance.
The gravitational potentials are related to each other (Eq.~\ref{eqn:phiE} and~\ref{eqn:psiE}) and
via the Poisson equation (Eq.~\ref{eqn:PphiE}) to the matter overdensity. The ISW effect is therefore proportional to
the quantity $d(D_{+}(a)/a)/da$, where $D_{+}(a)$ is the growth factor. Here we are in particular interested in the late
ISW effect because it is affected by the dark energy dynamics.

The Rees-Sciama (RS) effect is similar to the ISW, but includes non-linear evolution of the gravitational potentials, 
which we include to second order, following \cite{Schaefer2008}. The ISW effect depends on the time derivative of the 
gravitational potential $\Phi$ that, via Poisson's equation is related to the overdensity $\delta$:
$\nabla^2\Phi\propto\delta$. It is therefore possible to replace the gravitational potential with the overdensity itself
simply inverting Poisson's equation: $\Phi\propto\triangle^{-1}\delta$. Expanding the overdensity as
$\delta=D_{+}(a)\delta^{(1)}+D_{+}^2(a)\delta^{2}$ we obtain the desired expression for the RS effect.

The ISW and RS effects are shown in Fig.~\ref{fig:isw_rs} where, for clarity, we present differences between the 
cosmological models we considered and the $\Lambda$CDM model. The upper panel shows the ISW effect, the lower panel 
the RS effect. 
The largest differences between models occur at $z\approx 1$ and are of the order of $10\%$. As expected, the largest 
differences arise for the different couplings, while the non-minimally coupled dark energy models are all very 
similar to each other.

\begin{figure}
 \includegraphics[angle=-90,width=0.95\hsize]{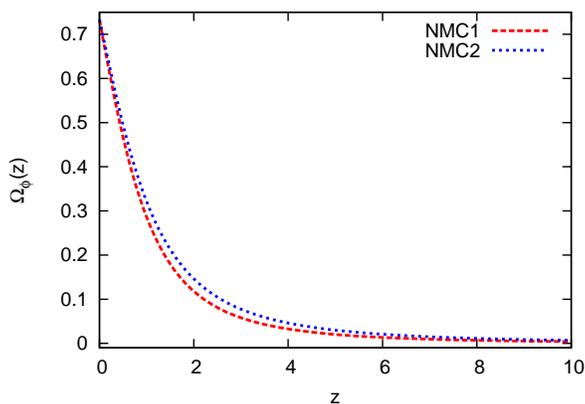}
 \caption{Amount of dark energy for the non-minimally coupled models as a function of redshift $z$. Line styles
and colours are as in Fig.~\ref{fig:wF}.}
 \label{fig:EDE}
\end{figure}

As we can see from Fig.~\ref{fig:isw_rs}, at early times when the dark energy contribution is negligible, all the 
models approximate the EdS model. We will therefore have that $D_+(a)\propto a$, hence the ISW vanishes while the RS
effect approaches an asymptotic value. This might appear surprising since non-minimally coupled models are characterized
by a non-negligible amount of dark energy at early times. However, for the models we consider, the coupling constant is
very small and the amount of dark energy at early times is negligible, as it is evident from Fig.~\ref{fig:EDE}.

\subsection{Spherical collapse}\label{sect:spcstt}
We next summarise theoretical arguments required to evaluate the spherical collapse parameters $\delta_{\rm c}$ 
(the linear evolution overdensity parameter) and $\Delta_{\rm V}$ (the virial overdensity parameter).

In the spherical collapse model, objects forming under gravitational collapse of matter over-densities
are assumed to be non-rotating and spherical. Even though this is clearly a crude assumption, since cosmic 
structures originate from the primordial seeds are triaxial and rotating \citep{Shaw2006,Bett2007}, the model provides 
predictions that can reproduce the results of numerical simulations quite well.
Spherical collapse has been analysed in the literature very extensively \citep[see, e.g.][]{Bernardeau1994, Ohta2003, 
Ohta2004, Mota2004, Nunes2006, Abramo2007, Basilakos2009, Pace2010, Basilakos2010, Wintergerst2010a}; 
to study perturbations in non-minimally coupled models we will follow closely
\cite{EspositoFarese2001,Perrotta2004,Acquaviva2004,Pettorino2008}.

In order to derive the differential equation describing the time evolution of the linear overdensity factor, we can 
simply repeat the derivation described above (Sect.~\ref{sect:pstt}), taking into account the full non-linearity of the 
continuity and Euler equations.
Doing so, the continuity and Euler equations read
\begin{eqnarray}
 \dot{\delta}+(1+\delta)\vec{\nabla}_{\vec{x}}\cdot\vec{u} & = & 0\label{eqn:pconNL}\\
 \frac{\partial\vec{u}}{\partial t}+2H\vec{u}+(\vec{u}\cdot\nabla_{\vec{x}})\vec{u}+\frac{1}{a^2}\nabla_{\vec{x}}\psi & 
= & 0\label{eqn:peulNL}\;.
\end{eqnarray}
We take the time derivative of Eq.~(\ref{eqn:pconNL}) and inserting into it the divergence of Eq.~(\ref{eqn:peulNL}),
with the help of the Poisson equation we obtain an exact second order non-linear differential equation describing the
evolution of matter perturbations,
\begin{equation}\label{eqn:nldelta}
 \ddot{\delta}+2H\dot{\delta}-\frac{4}{3}\frac{\dot{\delta}^2}{1+\delta}-
 4\pi G_{\mathrm{eff}}\bar{\rho}_{\mathrm{m}}\delta(1+\delta)=0\;,
\end{equation}
where $G_{\rm eff}$ is given by Eq.~(\ref{eqn:Geff}).

This is the non-linear equation we will use to infer the time evolution of the linear overdensity parameter
$\delta_{\mathrm{c}}$. Its linearised version is
\begin{equation}\label{eqn:ldelta}
\ddot{\delta}+2H\dot{\delta}-4\pi G_{\mathrm{eff}}\bar{\rho}_{\mathrm{m}}\delta=0\;,
\end{equation}
reproducing the classical result, but with $G\rightarrow G_{\mathrm{eff}}$.
It is worth pointing out that the correct linear growth has to be obtained from the non-linear equation for matter 
perturbations, Eq.~(\ref{eqn:nldelta}). (For a more complete discussion on this point, we refer to
\cite{Wintergerst2010a}.)
In order to evaluate the linear overdensity parameter $\delta_{\rm c}$, we use Eq.~(\ref{eqn:nldelta}) to find the 
initial conditions $\delta_{\rm i}$ and $\dot{\delta}_{\rm i}$ such that $\delta$ diverges at the chosen time of
collapse. Once the two initial conditions are found, we evolve them with the linear Eq.~(\ref{eqn:ldelta}), and its
density contrast at the collapse time gives $\delta_{\rm c}$.

\begin{figure}
\includegraphics[angle=-90,width=\hsize]{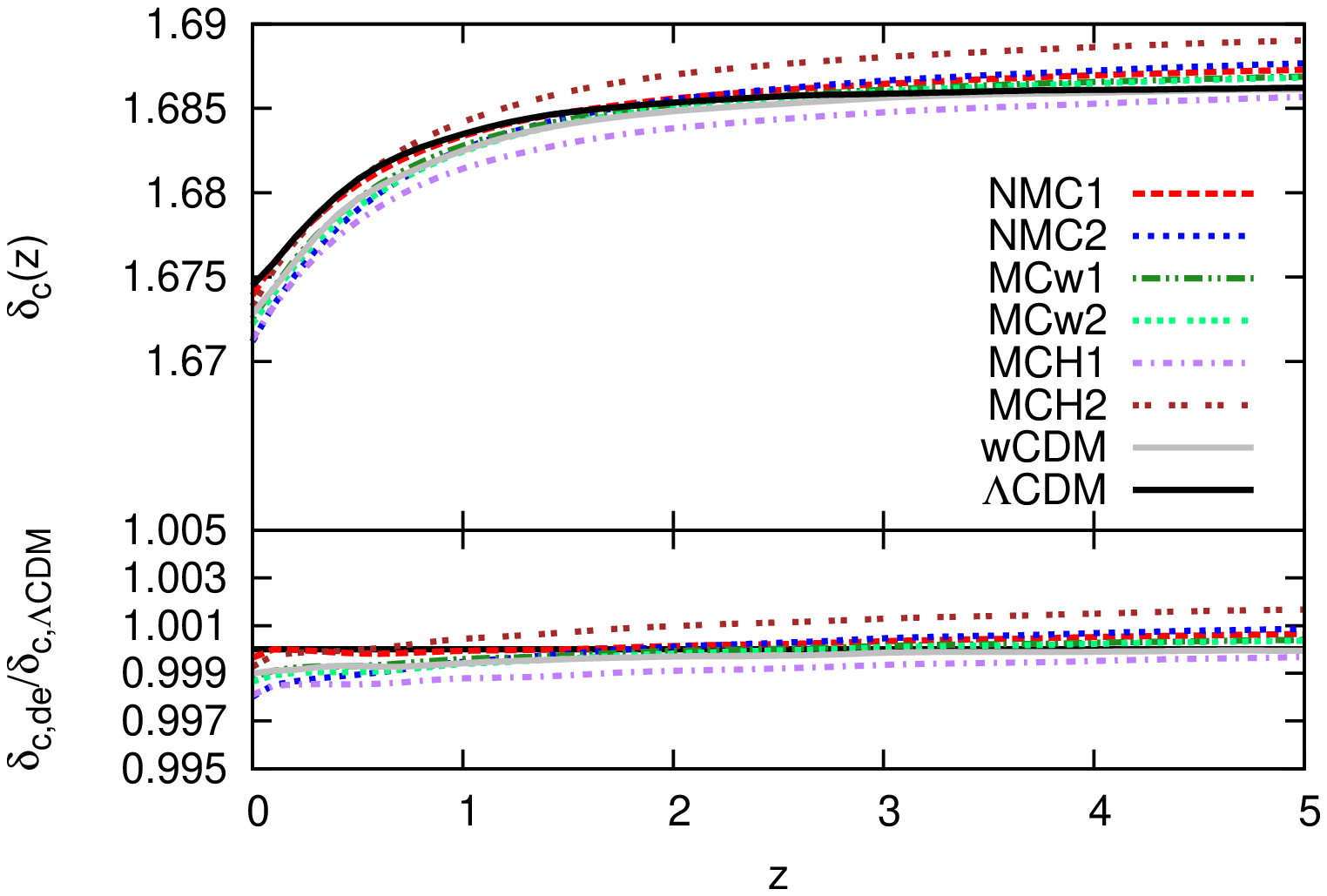}
\includegraphics[angle=-90,width=\hsize]{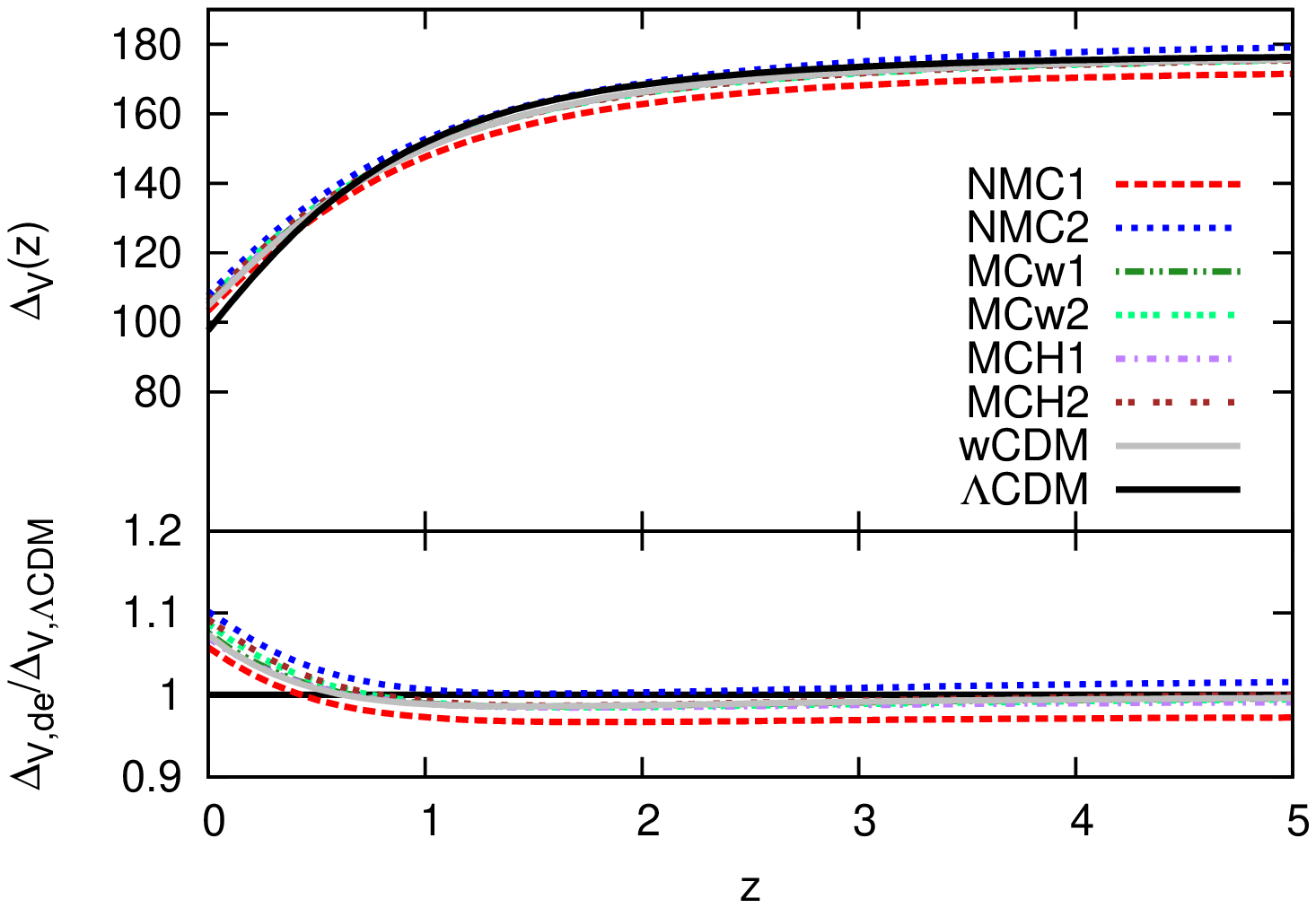}
\caption{The redshift evolution of the linear overdensity parameter $\delta_{\mathrm{c}}$ (top panel) and of the virial 
overdensity $\Delta_{\mathrm{V}}$ (bottom panel) for the models here considered. Each panel consists of two insets: the
upper one shows the absolute values of the quantities analysed, while the lower one the ratio between the dark energy
models and the reference $\Lambda$CDM one. Line types and colours are as in Fig.~\ref{fig:gf}.}
 \label{fig:dcdv}
\end{figure}

Our main results are presented in Fig.~\ref{fig:dcdv}. In the upper panel we show the time evolution of $\delta_{\rm c}$
while in the lower panel we present the time evolution of $\Delta_{\rm V}$.
We see that differences in $\delta_{\rm c}$ are very small, much below $1\%$ at $z=0$, while at high redshifts all 
the models converge to the Einstein-de Sitter (EdS) result. This is due to the stringent solar system constraints which
require a very weak coupling between the scalar field and the Ricci scalar. As seen in the lower panel of
Fig.~\ref{fig:wF}, gravity changes rapidly at low redshifts (where we expect the highest differences), while at high
redshifts the gravitational constant $G$ is practically constant and differs from the usual value by $\approx\pm 2\%$.
We also notice that models with a lower (higher) growth factor also have a smaller (larger) $\delta_{\rm c}$. The
NMC1 model is virtually indistinguishable from the $\Lambda$CDM model, while the largest differences appear for models
with negative coupling (NMC2). 
Model MCH2 shows lower values for the linear overdensity parameter $\delta_{\rm c}$ with respect to the $\Lambda$CDM
model also at high redshifts. We checked that this is not the case at high redshifts, where it is expected to approach 
the behaviour of an EdS model.

In the lower panel of Fig.~\ref{fig:dcdv}, we present results for the virial overdensity $\Delta_{\rm V}$. 
{ The virial overdensity is related to the non-linear evolution of the spherical overdensity. Given the turn-around 
scale factor $a_{\mathrm{ta}}$, when the radius of the collapsing sphere reaches its maximum value and
starts shrinking, the virial overdensity is defined as $\Delta_{\mathrm{V}}=\delta_{\mathrm{nl}}+1=\zeta(x/y)^3$, where
$x=a/a_{\mathrm{ta}}$ is the normalised scale factor and $y$ is the radius of the sphere normalised to its value at the
turn-around. For details on how to evaluate $\Delta_{\rm V}$ and the radius of the sphere $y$, we refer the reader to 
\cite{Pace2010}.}
The differences between the models studied here are also small, at most $10-15\%$, once again largest at low redshifts.
It is important to notice that at high redshifts, when naively we would expect to recover the result for the
$\Lambda$CDM model, this does not happen for the models with strongest absolute coupling value (models NMC1 and NMC2);
the differences are small in these cases, of order $2-3\%$, but still appreciable.
This kind of behaviour is expected, since $\Delta_{\mathrm{V}}$ is related to the solution of the non-linear equation 
for overdensities (Eq.~\ref{eqn:nldelta}) and we expect that the models will strongly differ from each other at the
non-linear level.

\begin{figure}
 \includegraphics[angle=-90,width=0.9\hsize]{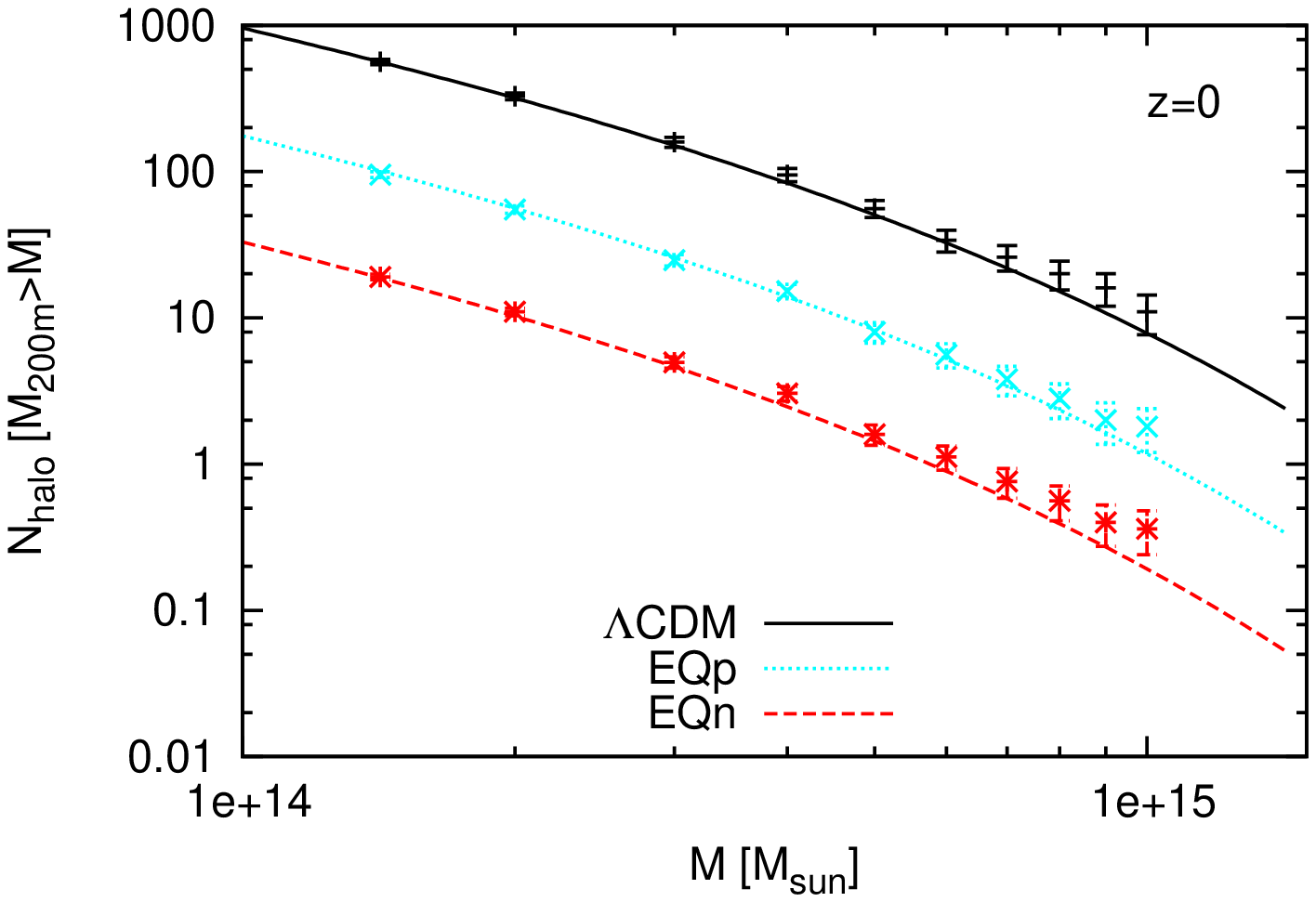}
 \includegraphics[angle=-90,width=0.9\hsize]{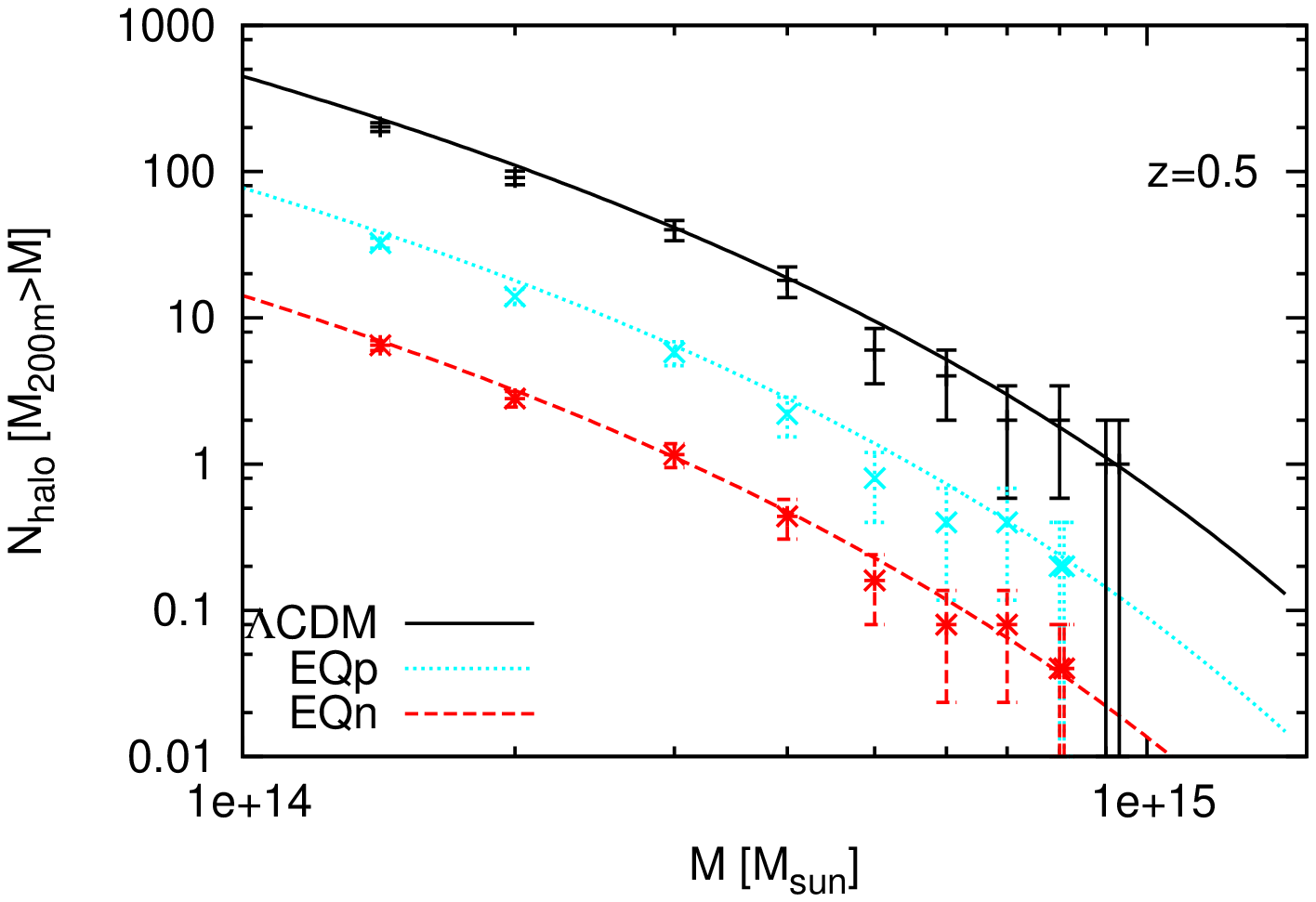}
 \includegraphics[angle=-90,width=0.9\hsize]{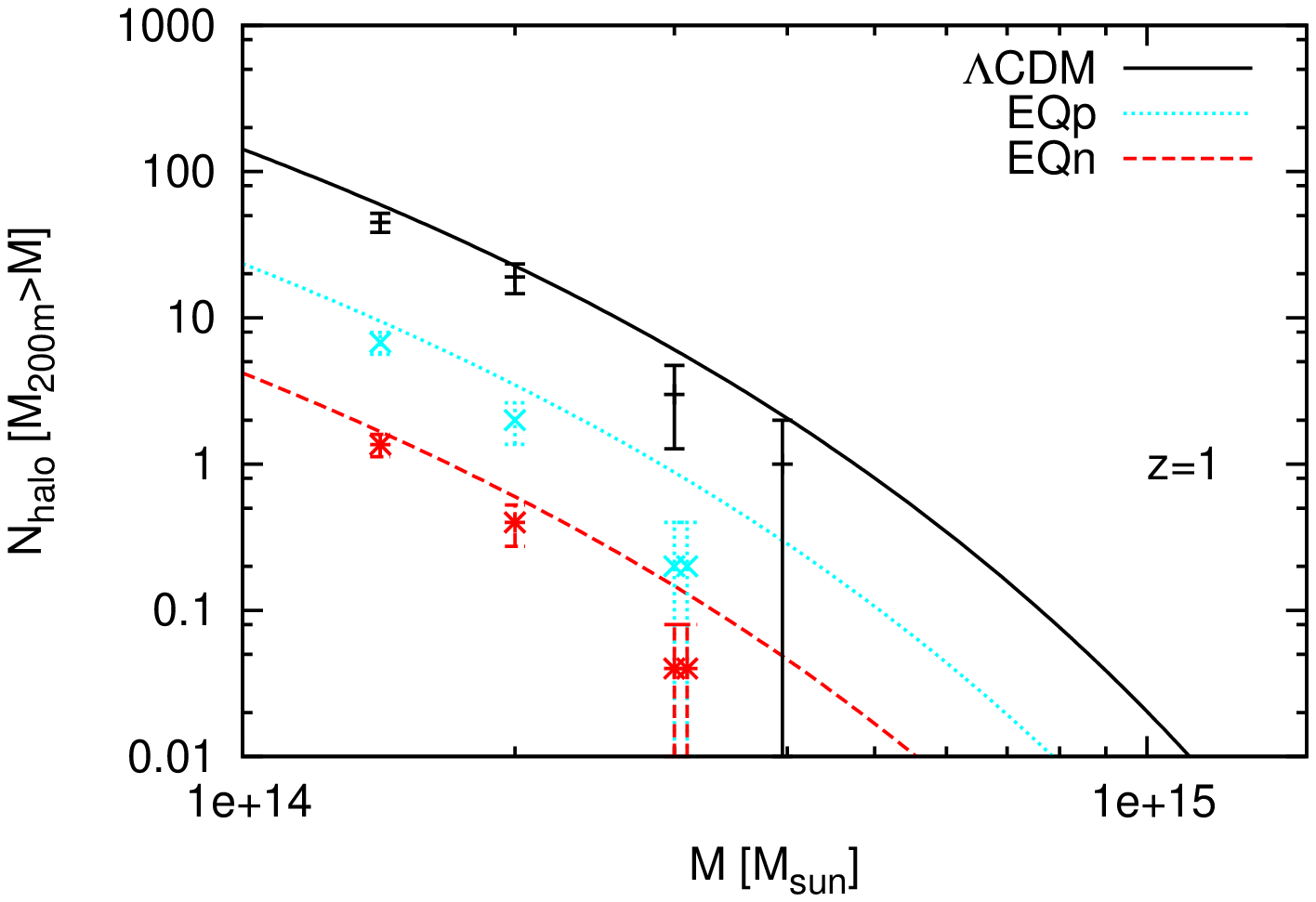}
 \caption{Comparison between the theoretical cumulative mass function and the results from N-body simulations. Black 
solid line and pluses represent the $\Lambda$CDM model, the cyan dotted line and crosses the EQp model (named NMC1 in 
this work) and red dashed line and stars the EQn model (named NMC2 in this work). Models EQp and EQn are scaled of a 
factor 5 and 25 respectively, for visualization purposes. Shown from top to bottom are comparisons at different 
redshifts: $z=0$, $z=0.5$ and $z=1$.}
 \label{fig:mfComp}
\end{figure}

Here we have assumed that the traditional recipes available in literature to evaluate the virial overdensity are still
valid (\cite{Wang1998}). This should be valid here, since we assume the scalar field only modifies the background; 
however, this might not necessarily be the case if perturbations in the scalar field are accounted for.  

It is interesting to notice that both at linear and non-linear level it is possible to see the effect of a
time-dependent gravitational constant. In particular, for the linear overdensity parameter $\delta_{c}$, while all the
models show similar values, the minimally coupled models, except for the wCDM model, have a higher value with respect to
the non-minimally coupled ones. The non-minimally coupled models have a very distinct signature: the lower is the value
of the coupling constant, the lower is the linear overdensity parameter. A similar argument, albeit with reversed
conclusions, applies to the non-linear virial overdensity.

The linear overdensity parameter $\delta_{\rm c}$ is not directly observable, but it is an important
quantity entering into the mass function. This subject will be discussed in the following section.
The virial overdensity is instead related to the definition of observed clusters in order to define virial mass and
virial radius. As seen before in the lower panel of Fig.~\ref{fig:dcdv}, differences are small ($<20\%$),
therefore using the $\Lambda$CDM value will not result in a big error on the halo definition.

\subsection{Mass function}\label{sect:mf}
We next discuss the mass function, which describes the number of collapsed objects of a given
mass that are formed at a given time in a unit volume. The mass function depends crucially on two factors, the linear
growth factor $D_{+}(a)$ and the linear overdensity parameter $\delta_{\mathrm{c}}$. Since these quantities, or more
precisely their ratio, appear quadratically in an exponential term, small deviations from the fiducial model can give
rise to huge differences in the mass function. While $\delta_{\mathrm{c}}$ is not an observable, the mass function, or
its integral over the mass, can be directly observed using large cosmological surveys, once the survey selection
functions are taken into account.

Another important ingredient for the mass function is the variance, defined via the relation
\begin{equation}
\sigma^{2}_{\mathrm{M}}=\frac{1}{2\pi^{2}}\int_0^{+\infty}k^2T^2(k)W^2_R(k)P_0(k)dk\;,
\label{eqn:sigma}
\end{equation}
where $P_0(k)$ represents the primordial matter power spectrum, $T(k)$ is the matter transfer function, and $W_R(k)$ is
the Fourier transform of the real space top-hat window function. Since quintessence models differ slightly from the
fiducial $\Lambda$CDM model \citep[see e.g.][]{Ma1999}, for simplicity we assume that all the models have the same
power spectrum shape, therefore the only difference will be in the spectrum normalization. For the different 
values adopted, we refer to Table~\ref{tab:params}. To evaluate the mass function, we use the expression derived by
\cite{Sheth1999}.

To validate our work, we compare our theoretical predictions for the cumulative mass function with the simulation 
results by \cite{DeBoni2011} at the same redshifts presented in their work, namely $z=0, 0.5, 1$. In
Fig.~\ref{fig:mfComp}, we show the total number of objects in the simulated cube compared to the theoretical
predictions. (For presentation purposes, we scaled the cumulative mass function of the model NMC1 and NMC2 by a factor
of 5 and 25, respectively.) As it is clearly seen, at $z=0$ (upper panel) we have a very good agreement between the
theoretical predictions and the numerical results up to $5-6\times 10^{14}~M_{\odot}$, while for higher masses
deviations are noticeable. This is expected, because in the simulations there are only very few objects in those mass
bins, due to the fact that the simulated box size is only 300~Mpc/h. The error bars, as can be seen in Fig.~6 in
\cite{DeBoni2011}, are quite large and our theoretical expectations are well within the $1-\sigma$ error bar. At $z=0.5$
(middle panel) the agreement between the theoretical predictions and the numerical mass function is still good over all
the mass range available from the simulations. At $z=1$ (lower panel), the agreement becomes substantially worse,
especially for the two non-minimally coupled models. This is due to the lack of objects at that redshifts in the
simulated volume; for the more numerous lower mass objects, up to $2\times 10^{14}~M_{\odot}$, the agreement between
simulations and analytic predictions remains good.

\begin{figure*}
\includegraphics[angle=-90,width=0.45\hsize]{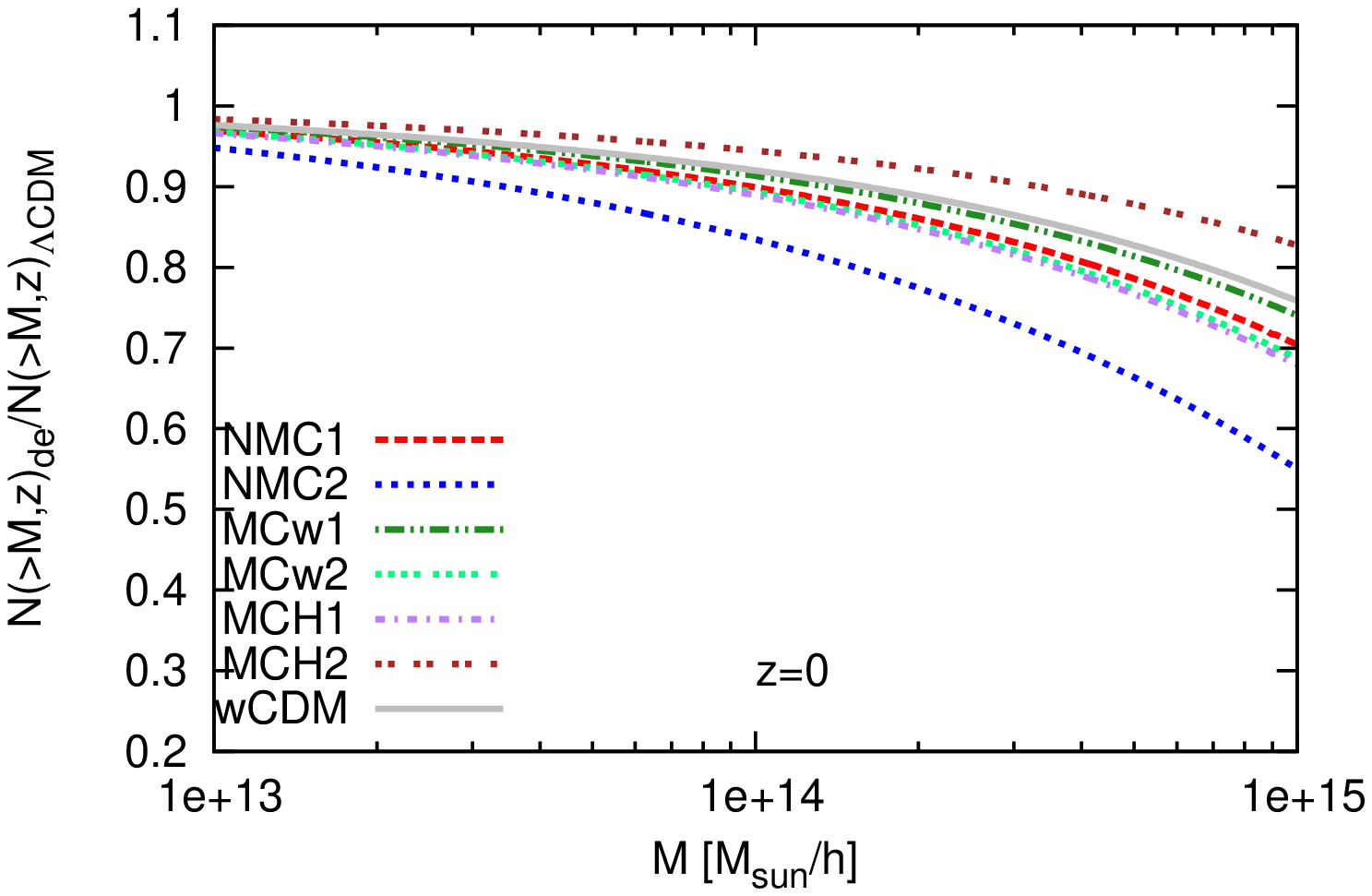}
\includegraphics[angle=-90,width=0.45\hsize]{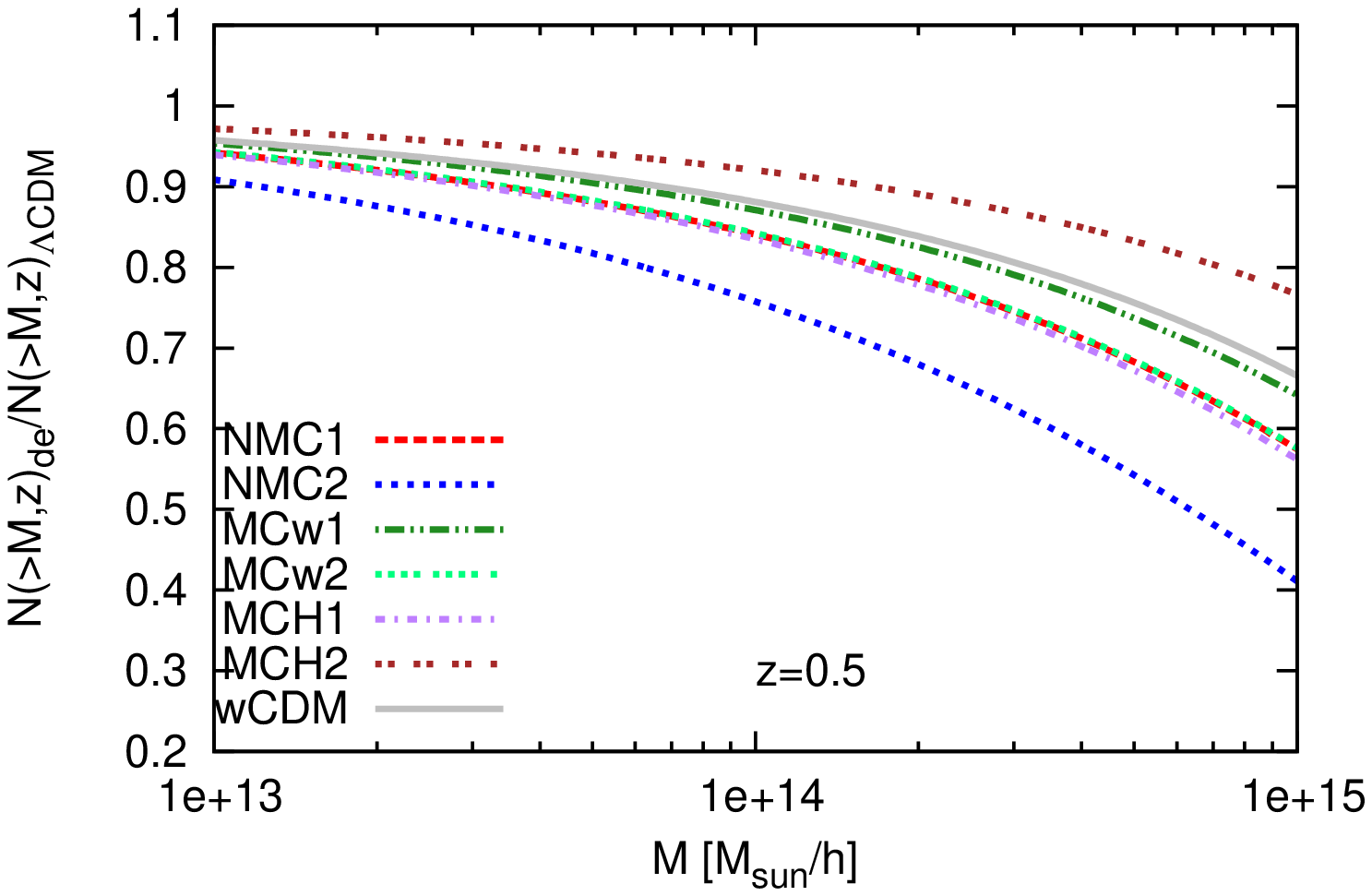}
\includegraphics[angle=-90,width=0.45\hsize]{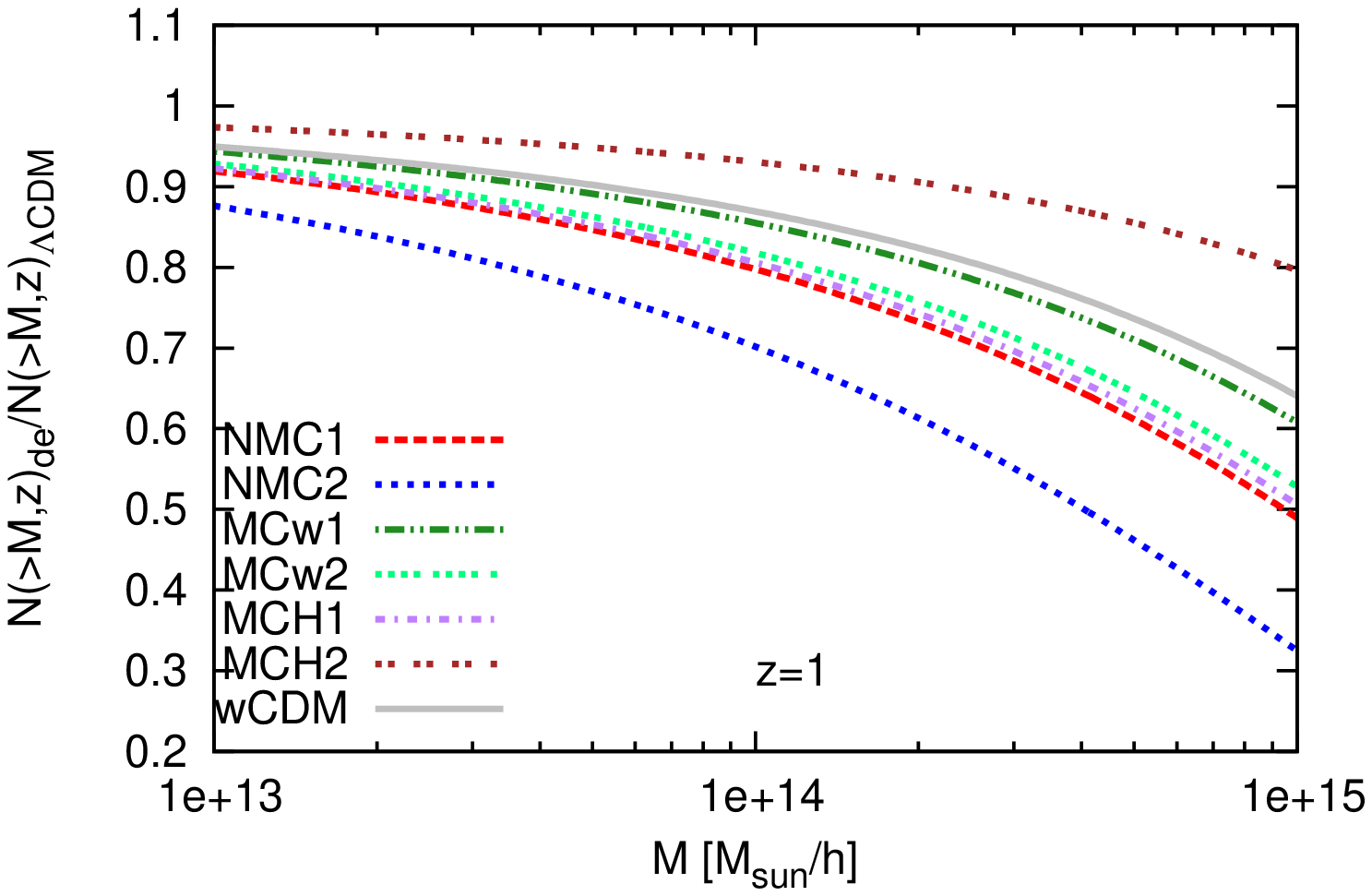}
\includegraphics[angle=-90,width=0.45\hsize]{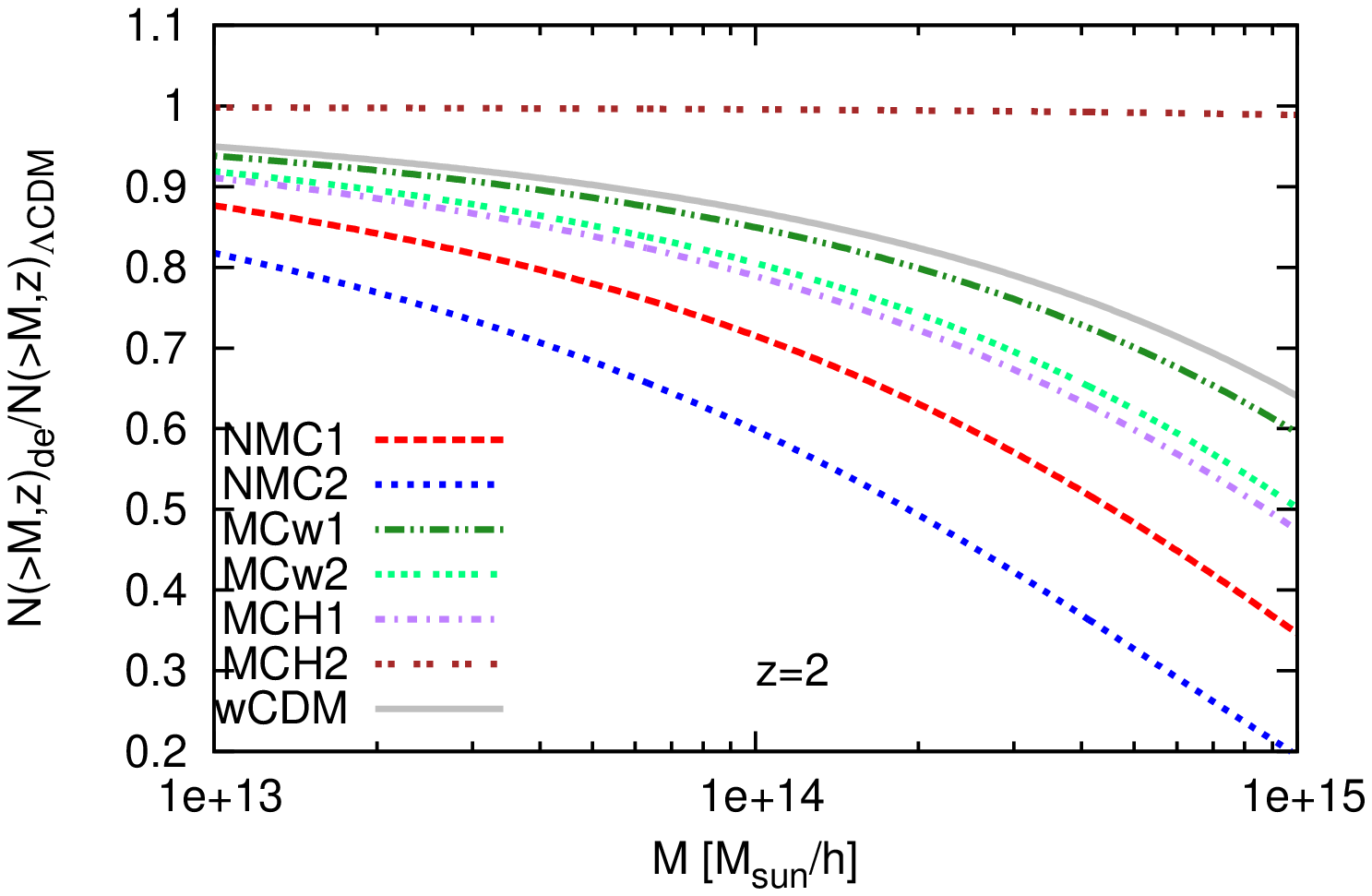}
\caption{Cumulative comoving number density of objects with mass exceeding $M$ at four different redshifts. Shown are 
ratios with respect to the reference $\Lambda$CDM model. Upper curves show the ratio with a normalization such that 
all the models have the same number of structure of the $\Lambda$CDM model at $z=0$, while lower curves have the same
amplitude of fluctuations at early times. Line types and colours are as in Fig.~\ref{fig:gf}.}
\label{fig:mf}
\end{figure*}

In Fig.~\ref{fig:mf} we show the ratio of the cumulative mass function for the dark energy models analysed with 
respect to the fiducial $\Lambda$CDM model. We evaluated the cumulative mass function, defined as the comoving number 
density of objects with mass exceeding $M$ at different redshifts, at four different redshifts, namely $z=0$, $z=0.5$,
$z=1$ and $z=2$.

By $z=0$ the models have substantial differences from the $\Lambda$CDM model, in particular they all show fewer structures. 
As expected, largest differences occur in the high mass tail, since rare events are affected more by changes in the growth 
of structure. Similar differences should appear in the void statistics. 
At $z=0$, the differences range from $10\%$ to $15\%$ for objects of mass $M\approx 10^{14}~M_{\odot}/h$ up to $40\%$ 
for very massive objects $M\approx 10^{15}~M_{\odot}/h$. At higher redshifts, the differences are even larger, in 
particular, at $z=2$ the model NMC2 has about $20\%$ of the number of very massive objects compared to that seen in the
$\Lambda$CDM model. Unfortunately at such high redshifts, the number of such massive clusters is so low that even large
differences are difficult to observe unless a very large volume of space is observed.

\begin{figure}
\includegraphics[angle=-90,width=0.85\hsize]{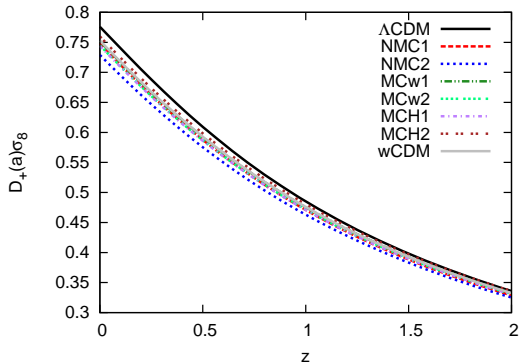}
\caption{Time evolution of the variance $\sigma_8$ for the different models studied. The square of this quantity is
important for the high mass tail of the mass function. Labels are like in Fig.~\ref{fig:gf}.}
\label{fig:gfs8}
\end{figure}

Differences between the non-minimally coupled models increase much faster than differences between the corresponding
minimally coupled models. In general, the minimally coupled models MCw1 and MCw2 show more
structures than the corresponding non-minimally coupled models NMC1 and NMC2. The wCDM model, with constant equation 
of state, is one of the closest to the $\Lambda$CDM predictions. This shows how important the evolution of the dark
energy equation-of-state parameter is for the mass function.
As we are normalising to the amplitude at early times, naively one might expect that the agreement with the $\Lambda$CDM model 
would be best at high 
redshifts, and in fact this is true for the MCH1 model.  However, at higher redshifts one is looking further into the tails of 
the distribution for a fixed mass, making the 
mass function more sensitive to small changes in the growth of structure to that time.
Due to the variation of the gravitational constant, the differences for the growth factor are higher for the non-minimally
coupled models, therefore the product $D_{+}(z)\sigma_8$ will be equal to the $\Lambda$CDM one at much higher redshifts. 
This is shown in Fig.~\ref{fig:gfs8} where we show the the product $D_{+}(z)\sigma_8$ for the different models studied here 
for the redshift interval $0\leqslant z\leqslant2$.
At higher redshifts massive objects are rare, therefore a small variation in the quantities related to structure formation
(growth factor and linear overdensity parameter) amplifies relative differences.

Many previous studies have dealt with alternative formulations of the halo mass function, based on fitting formulas of
the numerical mass function in the N-body simulations and assuming as fundamental variable the variance of the linear
matter power spectrum $\sigma_{\rm M}$ defined in Eq.~\ref{eqn:sigma}, \citep[see
e.g.][]{Jenkins2001,Reed2003,Warren2006,Reed2007,Crocce2010,Tinker2008,Courtin2011}. These formulations differ mainly 
on the high mass tail of the mass function and they could provide an higher fraction of massive objects.

In this work we adopt the prescription for the mass function following the work by \cite{Sheth1999}. The reason for
doing so is that the formulation of the mass function is motivated by the ellipsoidal collapse model and allowed us to
verify the validity of our calculations in the framework of the spherical collapse model. It has therefore a well
defined theoretical motivation, differently from the fitting formulas obtained for $\Lambda$CDM cosmologies, whose
validity for different cosmological models is not obvious. In particular, the numerical parameters used to evaluate the
mass function depend on the cosmological model studied and it is not obvious how to modify them for a new dark energy
model without having to determine them again from new simulations. This introduces the problem of the non-universality
of the mass function, deeply discussed in recent works by \cite{Lukic2007,Courtin2011,Reed2013}.

The interesting conclusion is that while a varying $G$ impacts on structure formation more strongly than a simple
non-minimally coupled dark energy model, one can infer differences between the models having the same background
history only at high redshifts.

We will see in Sect.~\ref{sect:lensing} how important this is when we study the cosmic shear power spectrum.

\subsection{Dark matter power spectrum}\label{sect:DMps}
A closely related statistic that can be used to study the dark matter clustering is the two-point correlation
function $\xi(r)$ and its Fourier transform, the matter power spectrum. On large scales, in the linear or mildly
non-linear regime, the power spectrum can be studied analytically, while for the fully non-linear regime it is necessary
to use either numerical N-body simulations or semi-analytic prescriptions fitted against simulations \citep[see
e.g.][]{Peacock1996,Smith2003}. Such approaches are limited in their validity by the scales that can be reached by
numerical simulations and on the models one can simulate. An alternative, physically-motivated approach is given by the
halo model developed by \cite{Ma2000a, Seljak2000} and others.

The halo model requires understanding in detail the mass function and the average dark matter density profile for a 
given model. Since these potentially depend on how the halo concentration changes with the coupling, it can be
difficult to calibrate for non-minimally coupled models. However, it may be hoped that most of the physics will be
captured in the $\Lambda$CDM model to first order, and in the following we will use power spectra obtained with the
prescription of the halofit, as outlined in \cite{Smith2003}. However, such uncertainties in the calibration must be
kept in mind here and in the following section which relates to the shear power spectrum (see Sect.~\ref{sect:lensing}).

\begin{figure}
\includegraphics[angle=-90,width=\hsize]{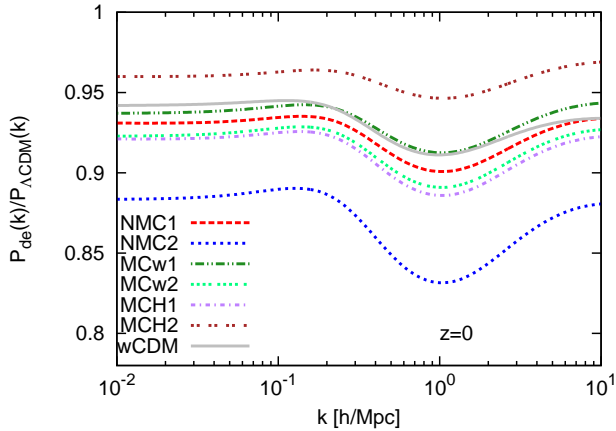}
\caption{Non-linear dark matter power spectrum for $z=0$ for the quintessence models here studied. Line types and 
colours are as in Fig.~\ref{fig:gf}.}
\label{fig:DMps}
\end{figure}

In Fig.~\ref{fig:DMps} we show the ratio of the dark matter power spectrum for the quintessence models to the same
quantity evaluated for the fiducial $\Lambda$CDM model as a function of the wave number. The matter power spectrum is
evaluated at $z=0$, using the CMB normalization described in Sect.~\ref{sect:models_params}, where we refer for the
exact normalization for each model. Using this normalisation, the models have different power at all scales, which on
linear scales results from integrating the different growth rates. As seen above, the largest differences arise for the
NMC2 model while the model differing least is MCH2, as its normalization is very close to the $\Lambda$CDM one.

The differences from the fiducial $\Lambda$CDM model are highest at the scale of $k\approx 1~h/$Mpc, where the 
dynamical dark energy matter power spectra show a dip (see also \cite{Ma2007}).
Since the power at all scales is significantly smaller than for the $\Lambda$CDM model, this results in a different 
scale where the power spectrum becomes non-linear and halofit corrections kick in. 
From a quantitative point of view, at large scales differences span a range between approximately $4\%$ and $12\%$ to
increase up to $16\%-17\%$ at $k\approx 1~h/$Mpc. The behaviour we found for the analytic power spectra is 
qualitatively in agreement with the analysis done by \cite{Fedeli2012} on the simulations presented by
\cite{DeBoni2011}.

Nonetheless we see that our results differ quantitatively from their analysis. In particular, comparing our results 
with their model labelled $DM_{0}$, we see that in our case the models differ more from what is seen in the simulations 
of approximately $3\%$ (see the lower panels in their Fig.~6). The major source of difference is related to the recipe
we adopted to evaluate the full non-linear matter power spectrum. From our figure, it is evident that the halofit
prescription can reproduce the nonlinear behaviour of the power spectrum up to few percent accuracy. We also notice that
the offset between the halofit prescription and the numerical simulations is roughly constant for the different models
analysed. Similarly to \cite{Fedeli2012}, we also find that the dip slightly changes position when a different
cosmological model is analysed. Moreover, as there speculated, the location of the dip is the same if the
background history of the models does not change. This is indeed the case for the couple of models NMC1 and MCH1 and
NMC2 and MCH2.

\subsection{Cosmic shear power spectrum}\label{sect:lensing}
\begin{figure}
\includegraphics[angle=-90,width=\hsize]{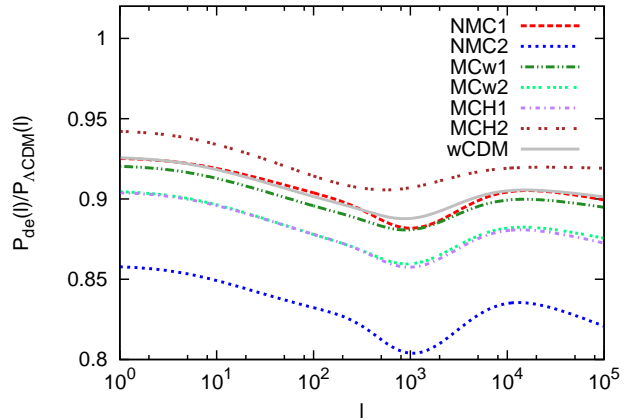}
\caption{Weak lensing power spectrum for the models analysed in this work. We present the ratio with respect to 
the $\Lambda$CDM shear power spectrum. Line types and colours are as in Fig.~\ref{fig:gf}.}
\label{fig:WLps}
\end{figure}

Gravitational lensing, where the images of background objects are distorted gravitationally, is an essential tool for 
understanding the distribution of dark matter. Measurements of weak lensing, where the distortions to the shapes of
objects are of order a few percent or less, are straightforward to predict and interpret for cosmological models. One
common weak lensing observable is the shear power spectrum, which is related to an integral along the line of sight of
the matter power spectrum. To evaluate the modifications to the form of the shear power spectrum, we follow the 
approach of \cite{Tsujikawa2008a} and \cite{Schimd2005}. Here we will just describe the most important steps in the
derivation of the final formula and we refer to their papers for more details. For a detailed analysis on the general
derivation of the lensing quantities for scalar-tensor theories, we refer to the work of \cite{Acquaviva2004}. Recently,
CMB lensing maps for a coupling in the Einstein frame that only involves dark matter were shown in \cite{Carbone2013}.

\begin{figure}
\includegraphics[angle=-90,width=\hsize]{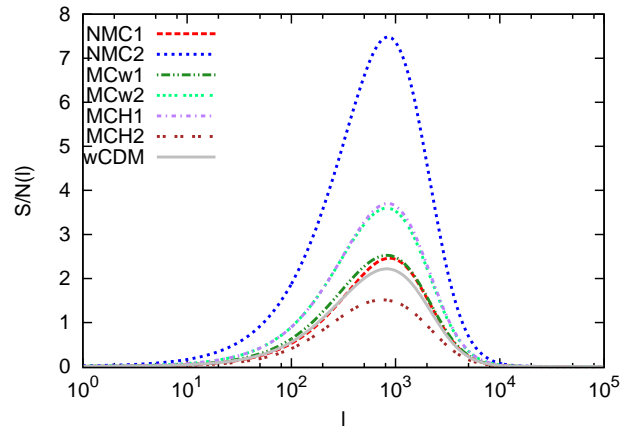}
\caption{The S/N ratio to distinguish between the concordance cosmology and each of the quintessence models
here considered as a function of the multipole. Line types and colours are as in Fig.~\ref{fig:gf}.}
\label{fig:SN}
\end{figure}

Starting from the perturbed metric
\begin{equation}
 ds^{2}=-(1+2\phi)dt^{2}+a^{2}(t)(1+2\psi)\delta_{ij}dx^{i}dx^{j}\nonumber\;,
\end{equation}
we can define the {\it deflecting potential}
\begin{equation}
\Phi_{\mathrm{wl}}=\phi+\psi\;,
\end{equation}
and the effective density field
\begin{equation}
\delta_{\mathrm{eff}}=\frac{a}{3H_{0}^{2}\Omega_{\mathrm{m},0}}k^{2}\Phi_{\mathrm{wl}}\;,
\end{equation}
where the relation between $\delta$ and $\delta_{\rm eff}$ is given by
\begin{equation}
\delta_{\mathrm{eff}}=\frac{\delta_{\mathrm{m}}}{F}\;.
\end{equation}
(Unlike in \cite{Tsujikawa2008a}, we do not have the term $F_{0}$ since in our case it is equal to one.)
The magnification matrix is defined as
\begin{equation}
A_{\mu\nu}=I_{\mu\nu}-\int_{0}^{\chi}\frac{\chi^{\prime}(\chi-\chi^{\prime})}{\chi}\partial_{\mu\nu}\Phi_{\mathrm{wl}}
d\chi^{\prime}\;,
\end{equation}
where $\chi$ is the comoving distance and $I$ is the identity matrix; from this, the effective convergence is given by 
\begin{equation}
\kappa=1-\frac{1}{2}\mathrm{tr}(A)\;.
\end{equation}
The shear power spectrum is related to the matter power spectrum by 
\begin{equation}\label{eqn:wlps}
P_\kappa(\ell)=\frac{9H^{4}_{0}\Omega^{2}_{\mathrm{m},0}}{4c^4}\int_{0}^{\chi_{\mathrm{H}}}\frac{W^{2}(\chi)}{a^{2}(\chi)
F^{2}(a)}P_{\delta_{\mathrm{m}}}\left[\frac{\ell}{f_{K}(\chi)},\chi\right]d\chi\;,
\end{equation}
where $f_{K}(\chi)$ is the comoving-angular diameter distance which depends on $K$, the spatial curvature of the 
Universe, and $P_{\delta_{\mathrm{m}}}$ is the matter power spectrum analysed in Sect.~\ref{sect:DMps}. The integral 
in the previous equation formally extends up to the horizon size $\chi_{\mathrm{H}}$, however since the number density
of sources (see below) drops to zero much before that, the integral can be effectively truncated at $z \sim 10$. The 
kernel $W(\chi)$ is an integral over the source redshift distribution which must be inferred from observations. 
In the following we will adopt the source redshift distribution derived by \cite{Fu2008} using data from the
Canada-France-Hawaii Telescope Legacy Survey (CFHTLS) and the parametrization for the non-linear matter power spectrum
given by \cite{Smith2003}, as discussed above.

In Fig.~\ref{fig:WLps} we show the ratio of the cosmic shear power spectrum for the models studied with respect to 
the prediction of the $\Lambda$CDM model. These follow to a large extent the trends observed in the matter power
spectrum (Fig.~\ref{fig:DMps}).
On large scales power spectra differ from $6\%$ to $13\%$ already, reflecting the normalization at high redshifts. The model
with the smallest differences is MCH2, while the model with the highest differences is the model NMC2. As expected, 
deviations from the fiducial model increase towards smaller angular scales, where the effects due to non-linearity are
more pronounced. The dip at $\ell\approx 10^3$ is a consequence of the analogous dip at $k\approx 1~h/Mpc$ seen in the 
power spectrum (see Fig.~\ref{fig:DMps}). We stress that these results are valid only for multipoles up to $\ell\sim
2000-3000$ since for smaller angular scales we would have to take into account baryonic physics.

To see how likely it is to observe the differences between the models considered, we look at the signal-to-noise (S/N)
ratio at a fixed multipole. 
The S/N ratio is defined as
\begin{equation}
\frac{S}{N}(\ell)=\left[\frac{P^\mathrm{DE}_\kappa(\ell)-P^{\Lambda\mathrm{CDM}}_\kappa(\ell)}{\Delta
P^{\Lambda\mathrm{CDM}}_\kappa(\ell)}\right]^2~,
\end{equation}
where $\Delta P^{\Lambda\mathrm{CDM}}_\kappa(\ell)$ is the Gaussian statistical error on the power spectrum in the 
framework of the concordance cosmology. According to \cite{Kaiser1992,Kaiser1998,Seljak1998,Huterer2002}, the latter 
can be evaluated approximately as
\begin{equation}
\Delta P^{\Lambda\mathrm{CDM}}_\kappa(\ell)=\sqrt{\frac{2}{(2\ell+1)\Delta\ell
f_{\mathrm{sky}}}}\left[P^{\Lambda\mathrm{CDM}}_\kappa(\ell)+\frac{\gamma^{2}}{\bar{n}_\mathrm{g}}\right]\;,
\end{equation}
where $\bar{n}_\mathrm{g}$ is the average surface number density of observed galaxies, $f_{\mathrm{sky}}$ is the 
fraction of sky area surveyed, and $\gamma$ represents the \emph{rms} intrinsic shape noise for the average galaxy. 
For practical purposes, we assume typical values for a future weak lensing survey and we set 
$\bar{n}=40$~arcmin$^{-2}$, $f_{\mathrm{sky}}=1/2$ and $\gamma=0.22$ \citep[see][]{Zhang2009}. As suggested by
\cite{Takada2007} and \cite{Takada2009} we use $\Delta\ell=1$.

\begin{figure*}
\includegraphics[angle=-90,width=0.9\hsize]{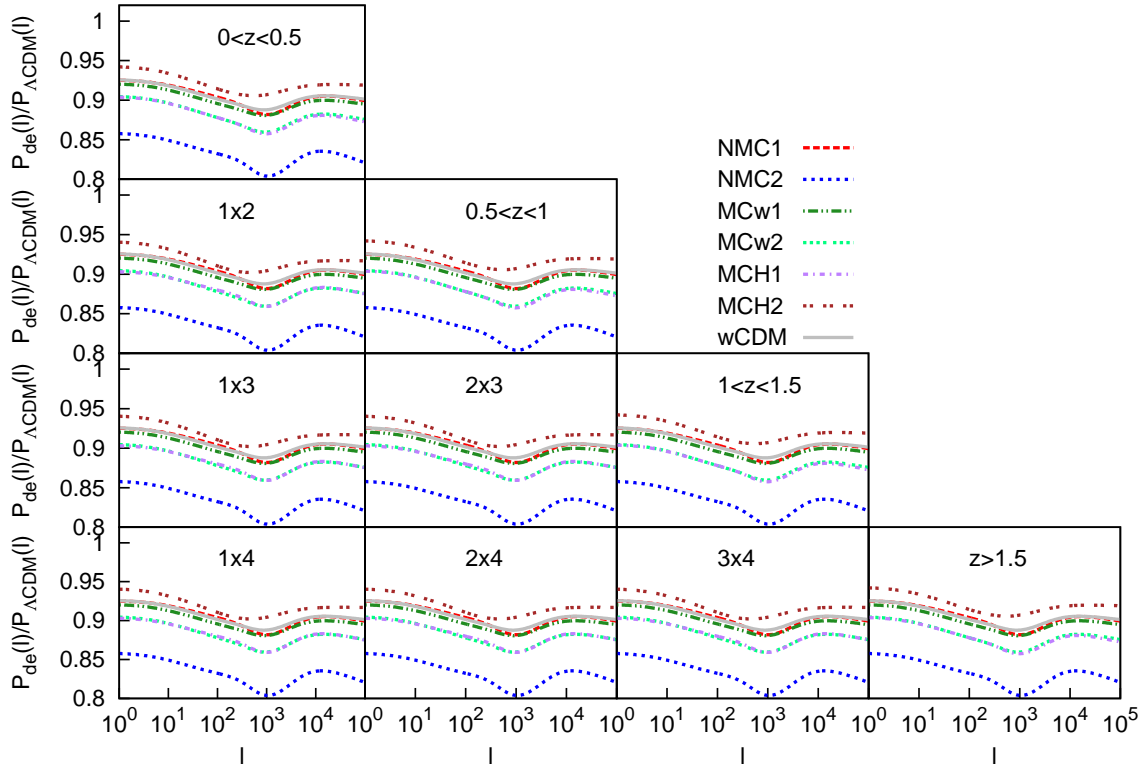}
\caption{The cosmic shear power spectrum for the quintessence models here analysed derived in a specific source 
redshift bin. We present the ratio with respect to the predictions of the fiducial $\Lambda$CDM model. The redshift 
bins are as follows: $0<z<0.5$, $0.5<z<1$, $1<z<1.5$ and $1.5<z<\infty$. The first three rows represent the
cross-correlation power spectra, while the last row shows the auto-correlation power spectra. We refer to the labels 
in the panels for the corresponding redshift bins. Line types and colours are as in Fig.~\ref{fig:gf}.}
\label{fig:tomography}
\end{figure*}

In Fig.~\ref{fig:SN} we show the $S/N$ ratio for the cosmic shear power spectrum as a function of the multipole 
$\ell$. We notice that at intermediate scales these models have a significant $S/N$ ratio and that it decreases 
very quickly for lower and higher multipoles; this is in agreement with what was seen by \cite{Fedeli2010} in the
context of non-Gaussianity in weak lensing and \cite{Pace2012} in the context of oscillating dark energy models. This
suggests that it will be very easy to differentiate the models via weak lensing techniques by summing just over few
multipoles.
Consistently with Fig.~\ref{fig:WLps}, the model with the highest $S/N$ ratio is the NMC2, which differs most from the
fiducial model.

A very important tool used to increase the power of cosmic shear is by using the tomography of lensing
\citep{Hu1999,Takada2004} and it consists in the subdivision of the sources in several bins, and computing the
shear power spectrum in each bin and the cross correlation between different redshift bins. More precisely, the cross 
power spectrum between two bins is
\begin{equation}
P_\kappa^{ij}(\ell)=\frac{9H^{4}_{0}\Omega_{\mathrm{m},0}^2}{4c^{4}}
\int_{0}^{\chi_{\mathrm{H}}}P\left(\frac{\ell}{f_{\mathrm{K}}(\chi)},\chi\right)\frac{W_{i}(\chi)W_{j}(\chi)}{a^{2}
(\chi)F^2(a)}d\chi\;,
\end{equation}
and now the redshift distribution has to be normalized to unity in each redshift bin, rather than the whole redshift 
range.

We considered four different bins, close to the maximum number that should give appreciable improvement given the 
broad lensing kernel \citep{Sun2009}, using the redshift intervals $[0,0.5]$, $[0.5,1]$, $[1,1.5]$ and $[1.5,\infty]$.
The results are shown in Fig.~\ref{fig:tomography}. In the bottom row we show the results for the ratio of the
auto-correlation power spectra while in the other panels we present the cross-correlated power spectra. The label
$m\times n$, where $m$ and $n$ run from one to four (total number of redshift bins), indicates the cross-correlation
between bins $m$ and $n$.

As it appears clear from Fig.~\ref{fig:tomography} we notice that the information carried by the auto-correlation 
power spectrum is very similar for all the models. We see similar behaviour in the cross-correlated shear power spectrum. 
All the minimally coupled models behave very similarly and differ from the $\Lambda$CDM model of $\approx 10\%$. 
As shown in Figs.~\ref{fig:WLps},~\ref{fig:SN} and~\ref{fig:tomography} differences between the dark energy models we consider
are quite pronounced. According to \cite{Beynon2012}, future lensing surveys as Euclid\footnote{www.euclid-ec.org}
\citep{Laureijs2011,Amendola2012a} will be able to differentiate models at the level of $2-3\%$. Since all the models analysed
here differ by the reference $\Lambda$CDM model for more than $5\%$ at all scales, we can safely conclude that future lensing
surveys will easily say whether these models will be compatible with the data or not.

\section{Conclusions}\label{sect:conc}
In this work we studied the structure growth and evolution of quintessence models, with particular emphasis on
non-minimally coupled models (scalar-tensor theories) where the effective gravitational constant $G$ changes in time.
We compared representative scalar-tensor models to standard GR models which are described by the same equation of 
state, and also to a simple constant equation of state model ($w=-0.9$). We also considered two additional minimally
coupled models where the background expansion is identical to the non-minimally coupled models. Our principle 
aim has been to isolate the influence of a varying gravitational constant $G$ on structure formation, extending recent 
numerical work on this subject \citep{DeBoni2011} by carrying out analytic predictions for the same models that were
previously simulated.

We studied several quantities, ranging from the linear analysis of the growth factor to the non-linearity of the mass 
function and of the weak lensing power spectrum. To validate our theoretical considerations, we compared our mass 
function to the one obtained directly from the N-body simulation (see Fig.~\ref{fig:mfComp}). We showed that an
analytical analysis of the linear growth factor ($D_{+}(a)$) and linear overdensity parameter $\delta_{\rm c}$ can
largely reproduce the numerical mass function over two orders of magnitude in mass.

A time-dependent $G$ has a greater impact on all the quantities we considered, as compared to a conventional dark 
energy model whose dark energy component possesses the same equation of state, but interestingly enough, differences
are mitigated, at least at the linear level, when the minimally coupled models have the same background expansion
history predicted in the framework of scalar-tensor theory.
The strength of gravity changes over time for scalar-tensor theories, adding one more 
degree of freedom to the standard general relativistic framework. In the models we considered $G$ varied up to $2\%$, 
leading to changes in background quantities at a similar level; however, at the perturbation level differences are 
amplified. For example, the growth factor can change up to $10\%$ and the range of variation is dictated by the models
with the most extreme coupling (NMC1 and NMC2).

Similar comparisons can be made for the critical linear density contrast for spherical collapse $\delta_{\rm c}(z)$ 
and the virial overdensity $\Delta_{\rm V}(z)$. These remain similar to the predictions for a $\Lambda$CDM model, 
differing from it by only few percent or less. As expected, the models converge to the prediction for an Einstein-de
Sitter universe at high redshifts, but the rate of convergence is model dependent and it is influenced by the amount of
dark energy at early times.

These small differences are amplified when looking at the mass function for rare objects; differences from the fiducial
model are large, of the order of $40\%$ for very massive objects already at $z=0$ and can be as large as $80\%$ at a
redshift $z=2$, where one is probing even rarer objects. Deviations from the $\Lambda$CDM model are generally amplified
if the gravity strength changes in time.

The dark matter power spectrum shows differences of the order of $10\%-15\%$ at most, particularly on mildly 
non-linear scales. On large scales differences are mostly due to integrated differences in the growth rate, ranging 
from $5\%$ to $10\%$. These conclusions are in qualitative agreement with those found by \citet{Fedeli2012} of the
analysis of N-body simulations and differ by only a few percent, showing that the usual recipes for the matter power
spectrum can reproduce these models reasonably well without further calibration. The small differences are due to
the fact that we assume that the assumptions used to build the halofit model are still valid in non-minimally coupled
models (see discussion of how the non-linear matter power spectrum was evaluated in their Sect.~4).

Finally, the effective convergence power spectrum is affected at the level of $\sim 10-15\%$ at intermediate/small
angular scales. Since the corresponding observations are in principle very sensitive, it will be possible to discriminate 
these models with future lensing surveys, such as Euclid. In particular, as discussed in \cite{Beynon2012} a precision of 
few percent can be reached. This implies that all the models could in principle be falsified if $\Lambda$CDM is the true 
cosmological model.

\section*{Acknowledgements}
The authors thank Cristiano De Boni for providing the numerical mass function of their N-body simulations and Mischa
Gerstenlauer, David Bacon and Emma Beynon for useful discussions. FP and RC are supported by STFC grant ST/H002774/1. 
VP is supported by Marie Curie IEF, Project DEMO. LM acknowledges financial contributions from contracts  ASI/INAF
I/023/12/0, by the PRIN-MIUR09 “Tracing the growth of structures in the Universe” and by PRIN MIUR 2010-2011 “The dark
Universe and the cosmic evolution of baryons: from current surveys to Euclid".

\bibliographystyle{mn2e}
\bibliography{nmcModel.bbl}

\label{lastpage}
\end{document}